\documentclass{article}
\usepackage{arxiv}

\usepackage[utf8]{inputenc} 
\usepackage[T1]{fontenc}    
\usepackage{hyperref}       
\usepackage{url}            
\usepackage{booktabs}       
\usepackage{microtype}      
\usepackage{lipsum}	    	
\usepackage{graphicx}
\usepackage{doi}
\usepackage[section]{placeins}
\usepackage{mathtools}
\usepackage{amsmath}
\usepackage{amssymb}
\usepackage[ruled,vlined]{algorithm2e}
\usepackage{lineno}
\usepackage[dvipsnames]{xcolor}
\usepackage{enumitem}
\usepackage{placeins}
\usepackage{multirow}
\usepackage[sort&compress,numbers]{natbib}
\usepackage{xcolor}                
\usepackage{caption} 
\usepackage{tikz}
\usetikzlibrary{
  positioning,
  arrows,
  shapes.geometric,
  fit,
  backgrounds,
  calc,
  decorations.pathreplacing,
  matrix
}
\definecolor{branchcolor}{RGB}{70,130,180}
\definecolor{trunkcolor}{RGB}{60,179,113}
\definecolor{outputcolor}{RGB}{255,165,0}
\definecolor{flowcolor}{RGB}{47,79,79}
\setlist[itemize]{noitemsep, topsep=0pt}
\usepackage{color}
\usepackage{soul}

\title{\textbf{Time Resolution Independent Operator Learning}}

\author{{\hspace{1mm}Diab W. Abueidda}\thanks{da3205@nyu.edu} \\
	Civil and Urban Engineering Department\\
	New York University Abu Dhabi\\
        National Center for Supercomputing Applications\\
	University of Illinois at Urbana-Champaign\\
        \And
        {\hspace{1mm}Mbebo Nonna} \\
        Computer Science Department\\
        New York University Abu Dhabi\\
	\And
        {\hspace{1mm}Panos Pantidis}\\
	Civil and Urban Engineering Department\\
	New York University Abu Dhabi\\
        \And
	{\hspace{1mm}Mostafa E. Mobasher}\thanks{mostafa.mobasher@nyu.edu}\\
	Civil and Urban Engineering Department\\
	New York University Abu Dhabi\\
}

\renewcommand{\shorttitle}{NCDE-DeepONet}

\begin{document}

\maketitle

\begin{abstract}
Accurately learning solution operators for time-dependent partial differential equations (PDEs) from sparse and irregular data remains a challenging task. Recurrent DeepONet extensions inherit the discrete-time limitations of sequence-to-sequence (seq2seq) RNN architectures, while neural-ODE surrogates cannot incorporate new inputs after initialization. We introduce \emph{NCDE‑DeepONet}, a continuous‑time operator network that embeds a Neural Controlled Differential Equation (NCDE) in the branch and augments the trunk with explicit space–time coordinates. The NCDE encodes an entire load history as the solution of a controlled ODE driven by a spline-interpolated input path, making the representation \emph{input-resolution-independent}: it encodes different input signal discretizations of the observed samples. The trunk then probes this latent path at arbitrary spatial locations and times, rendering the overall map \emph{output‑resolution independent}: predictions can be queried on meshes and time steps unseen during training without retraining or interpolation. Benchmarks on transient Poisson, elastodynamic, and thermoelastic problems confirm the robustness and accuracy of the framework, achieving almost instant solution prediction. These findings suggest that controlled dynamics provide a principled and efficient foundation for high-fidelity operator learning in transient mechanics.
\end{abstract}

\keywords{Neural differential equations \and Ordinary differential equations \and Operator learning \and Thermoelasticity \and Elastodynamics}

\section{Introduction} \label{intro}

Deep learning techniques for operator learning have enabled data-driven modeling of complex physical systems by approximating mappings between function spaces. Modeling transient mechanics problems often requires learning the relationship between temporal inputs and outputs. Traditional neural network approaches, such as multilayer perceptrons (MLPs), are not naturally suited to this task because they operate on fixed-size input-output pairs and lack an inherent mechanism for sequential dependence \cite{koric2021deep, wu2022recurrent, herrmann2023deep}. Recurrent neural networks (RNNs) introduce a hidden state to capture sequence history. Recurrent neural networks (RNNs) such as Long Short-Term Memory (LSTM) and Gated Recurrent Unit (GRU) have recently been applied to capture nonlinear, history-dependent behavior in solid mechanics \cite{koric2021deep, wu2022recurrent, herrmann2023deep, su2024thermodynamics, bonatti2022importance}. However, their discrete-time updates and training challenges (e.g., vanishing gradients and sensitivity to time-step size) can limit accuracy and stability \cite{kidger2020neural, ribeiro2020beyond, maia2023physically}. Sequence‑to‑sequence (seq2seq) surrogates show rigidity. Specifically, learned networks are locked to the temporal grid seen during training and cannot refine predictions at unseen times or accommodate irregularly sampled data, which severely limits their usefulness in transient mechanics applications \cite{wu2022recurrent, he2023machine, pantidis2024fenn}. Additionally, standard RNN-based surrogates for history-dependent material models have been shown to produce oscillatory or inconsistent results under perturbations in the input, failing to converge when embedded in numerical solvers \cite{wu2022recurrent, he2024incremental, he2023machine, bonatti2022importance}. These limitations motivate the search for modeling frameworks that can more naturally handle continuous-time dynamics and complex temporal dependencies.

Neural ordinary differential equations (NODEs) offer a promising paradigm for continuous-time modeling. A NODE treats the evolution of a neural network’s hidden state as an initial value problem defined by an ordinary differential equation, essentially providing a continuous-depth generalization of deep networks like residual networks \cite{chen2018neural, anumasa2022latent}. In a NODE, the hidden state $\boldsymbol{z}(t)$ is defined by a learned vector field—its detailed formulation is provided in Section \ref{NCDEs}—where the state evolves from an initial condition $\boldsymbol{z}(t_0)=\boldsymbol{z}_0$. This formulation ties the model’s forward pass to the solution of an ODE, which can yield advantages in parameter efficiency \cite{lee2021parameterized} and in capturing smooth dynamics \cite{rojas2021reduced, dutta2021neural, portwood2019turbulence}. However, a key drawback is that the trajectory of a NODE is entirely determined by its initial condition, with no mechanism to adjust the state based on new input information during the evolution \cite{kidger2020neural}. In other words, once the integration starts, the NODE cannot directly ingest time-varying external forcing or additional observations. This contrasts with an RNN, wherein the hidden state is updated at discrete time steps based on the current input $\boldsymbol{X}_n$ (a detailed formulation is provided in Section \ref{NCDEs}). The lack of a similar input injection in NODEs makes them insufficient for many transient mechanics problems—especially those driven by sequential loads or boundary conditions—unless one resorts to ad-hoc techniques.

Neural controlled differential equations (NCDEs) have been introduced as a continuous-time generalization of RNNs to overcome the above limitation \cite{kidger2020neural, kidger2022neural}. Just as NODEs are the continuous analog of residual networks, a Neural CDE serves as the continuous-time counterpart of an RNN \cite{kidger2020neural}. In an NCDE, the hidden state $\boldsymbol{z}(t)$ evolves according to a controlled ordinary differential equation driven by the input function. This construction provides a principled way to incorporate external time-dependent inputs into the model: the hidden state is continuously adjusted based on changes in the input, rather than solely on its initial value. As a result, NCDEs can naturally process incoming data that may be irregularly sampled or varying in time \cite{kidger2020neural, morrill2021neural}, while maintaining the smooth interpolation capabilities of ODE-based models \cite{kidger2020neural}. The NCDE approach essentially equips the network with a form of memory and the ability to handle complex temporal patterns, much like an RNN, but in a mathematically continuous framework \cite{morrill2021neural_b, bleistein2023generalization}. This makes NCDEs well-suited for capturing the causal, path-dependent behavior inherent in transient physical systems \cite{choi2022graph, jhin2023learnable}.

Deep operator networks (DeepONets) provide a complementary perspective by focusing on learning operators – mappings between function spaces – rather than simple point-to-point mapping \cite{lu2021learning, lu2021deepxde, abueidda2025deepokan}. DeepONet employs a dual-network architecture – a branch network to encode the input function and a trunk network to encode the query location (e.g., temporal or spatial coordinates) for the output, thereby effectively parametrizing the input and output function spaces. The network achieves this by combining the processed features from both networks through a Hadamard product operation, resulting in a flexible framework capable of handling various partial differential equations and functional relationships \cite{lu2022multifidelity, goswami2022physics, garg2023vb, kobayashi2024improved, cai2021deepm}. Notable advantages of DeepONet include its ability to maintain accuracy across different discretization schemes and its capacity to generalize to unseen input functions. Additionally, this construction of the DeepONet enables it to learn solution operators from relatively small datasets. Hence, DeepONet is becoming particularly valuable for applications in scientific computing and engineering simulations, where rapid and accurate solution approximations are essential \cite{goswami2023physics, kobayashi2024improved, lu2022comprehensive}.

Various extensions of the basic DeepONet framework replace the standard MLP-based branch and trunk networks with specialized architectures to better capture problem-specific structure. For instance, convolution-based DeepONets have been proposed (e.g., using CNNs in the branch net) to exploit local spatial correlations and translation invariance in input functions, thereby improving sample efficiency and generalization on high-dimensional fields \cite{ingebrand2025basis, kumar2024convolutional}. Additionally, to handle complex, spatially varying outputs, He et al. \cite{he2023novel} devised a DeepONet variant using a residual U-Net (ResUNet) as the trunk network, replacing the fully connected trunk, which enables the model to effectively encode intricate input geometries when predicting elastoplastic stress fields. This convolutional trunk significantly improved the network’s ability to learn from variable shapes. Similarly, to incorporate temporal dependencies, He et al. \cite{he2024sequential} introduced sequential DeepONet (S-DeepONet), which integrates a recurrent branch network for time-dependent input histories \cite{he2024sequential, he2024predictions}. Moreover, attention-based architectures, such as Transformers, have been explored as drop-in replacements for the MLP, with self-attention enabling the capture of long-range or global dependencies. Such transformer-augmented operator learners satisfy the universal approximation property and can even outperform standard DeepONets on challenging PDE benchmarks (albeit with a higher computational cost) \cite{shih2025transformers, yun2019transformers, zappala2024universal}. Another class of DeepONet variants integrates spectral or multi-scale representations into the branch or trunk. For example, POD-DeepONet projects the output onto a fixed PCA-based basis, effectively constraining the trunk network to a low-dimensional spectral subspace and achieving better accuracy on some problems \cite{kovachki2023neural}. Likewise, multi-scale DeepONet designs introduce networks capable of capturing high-frequency content (e.g., using Fourier features or wavelet bases) to mitigate the spectral bias of deep neural operators \cite{wang2025multi, tancik2020fourier, liu2024mitigating, wang2021eigenvector}. All these extensions share a common motivation: by endowing DeepONet with appropriate inductive biases (such as spatial locality, sequence modeling, and frequency-domain structure), one can significantly enhance its generalization ability and data efficiency while preserving its theoretical universality in approximating operators. 

Sequential Deep Operator Networks (S‑DeepONet) learn solution operators under time‑dependent loads by embedding a recurrent neural network (RNN) in the branch net \cite{he2024predictions, he2024sequential}. Although this architecture ingests complete loading histories, it suffers from two limitations: \textit{(i)} predictions are returned only at the final increment or a few user‑defined snapshots, and \textit{(ii)} the hidden state is updated at fixed discrete steps, so the learned operator is tied to the temporal resolution of the training data. In other words, once trained on a specific time step size, S‑DeepONet cannot easily refine its predictions at unseen intermediate times or accommodate irregularly sampled inputs, which limits its flexibility in transient mechanics applications. These constraints have spurred several extensions aimed at relaxing the time-step rigidity and improving long-horizon accuracy. For example, the Time-Integrator DeepONet (TI‑DeepONet) explicitly learns the system’s time derivative and embeds a numerical integrator (e.g., Runge-Kutta) into the model, preserving the Markovian dynamics and substantially mitigating error propagation over extended forecasts \cite{nayak2025ti}. Similarly, Time-Adaptive Operator Learning via Neural Taylor Expansion (TANTE) replaces the fixed-step rollout with a neural Taylor series expansion: at each step, the model learns higher-order temporal derivatives and a data-driven $\Delta t$ for adaptive stepping \cite{wu2025tante}. 

A separate challenge is the resolution dependency inherent in standard DeepONet, which requires input functions to be sampled on an identical, fixed sensor grid across all training samples \cite{tretiakov2025setonet, prasthofer2022variable}. Graph-based approaches have also been explored. For example, GraphDeepONet integrates a graph neural network with DeepONet to handle irregular spatial meshes and perform autoregressive time stepping, yielding robust accuracy on unstructured grids while enabling time extrapolation beyond the training interval \cite{cho2024graphdeeponet}. The Resolution Independent Neural Operator (RINO) takes a different route by eliminating the need for shared sensor locations \cite{bahmani2025resolution}. RINO learns continuous input and output dictionaries (basis functions parameterized as implicit neural representations) that project each function onto a finite-dimensional coefficient space. In effect, the operator learning task is reduced to mapping input basis coefficients to output basis coefficients, allowing the model to ingest arbitrarily sampled input/output data without retraining. This provides spatial resolution independence in both training and inference. However, RINO’s mapping occurs only between discrete snapshot pairs (e.g., initial and final states), with no continuously evolving latent state linking those snapshots Consequently, none of those mentioned above methods fully resolves S‑DeepONet’s limitations—no existing approach accepts irregularly sampled temporal data and produces outputs at arbitrary query times in a single framework. This gap motivates the search for new modeling paradigms (as pursued in our work) that can naturally handle continuous-time dynamics with both input- and output-resolution independence.

The proposed \textbf{NCDE‑DeepONet} closes this gap.  A Neural Controlled Differential Equation in the branch treats each load history as a continuous control signal, making the latent dynamics \emph{input‑resolution independent}: additional or irregular observations modify the trajectory without retraining. By injecting explicit space–time coordinates into the trunk, the network probes that trajectory at \emph{arbitrary} spatial points and instants, thereby conferring \emph{output-resolution independence}.  Hence, a model trained on a coarse or uneven mesh can be interrogated on a much finer—or entirely different—mesh during inference.  This continuous-time formulation offers a principled and scalable approach to operator learning for transient mechanics, although discrete-time surrogates may still excel in narrowly defined regimes.

The remainder of this paper is organized as follows. Section \ref{methods} provides the necessary background on recurrent neural networks, neural differential equations (NODEs and NCDEs), and deep operator networks. Section \ref{ncdedeeponet} introduces the NCDE-DeepONet architecture, detailing how the Neural CDE branch and spatiotemporal trunk work together to achieve input- and output-resolution independence, along with training considerations using the adjoint method. Section \ref{results} demonstrates the framework's effectiveness through three benchmark problems—transient Poisson, elastodynamics, and transient fully coupled thermoelasticity—with comparisons to GRU-DeepONet and an analysis of the model's unique interpolation capabilities. Finally, Section \ref{conclu} concludes with a summary of contributions and directions for future research.

\section{Methods}\label{methods}

The methods surveyed in this section are organized around their treatment of temporal data.  We begin with \textit{recurrent neural networks} (RNNs), the canonical discrete‑time sequence models in machine learning.  An RNN receives an input function observed at a finite set of time instants $\{t_n\}_{n=0}^{N}$, propagates a latent \emph{hidden state} that accumulates all past information, and generates an output sequence (or a final summary) of the same temporal structure. This recursive hidden state update is the architectural hallmark that enables RNNs to capture temporal dependencies, but it also ties the model to the fixed time grid used during training.  Building on this baseline, Section \ref{NCDEs} reinterprets the recurrence in the continuous‑time limit, leading to neural ordinary and controlled differential equations, while Section \ref{DeepONet} reviews deep operator networks that lift pointwise mappings to mappings between function spaces. Together, these concepts form the foundation for the NCDE‑DeepONet architecture introduced later in Section \ref{ncdedeeponet}.

\subsection{Recurrent Neural Networks (RNNs)}\label{rnn}
RNNs update the hidden state at discrete time intervals. The update rule is expressed as
\begin{equation}\label{rnn_equation}
\boldsymbol{h}_{n+1} = \hat{f}(\boldsymbol{h}_n, \boldsymbol{X}_n; \boldsymbol{\theta}),
\end{equation}
where:
\begin{itemize}
    \item $\boldsymbol{h}_n \in \mathbb{R}^d$ is the hidden state at the $n$th time step,
    \item $\boldsymbol{X}_n$ denotes the input at time step $n$,
    \item $\hat{f}$ is a nonlinear function that updates the state based on the current hidden state and input,
    \item $\boldsymbol{\theta}$ represents the learnable parameters.
\end{itemize}
This discrete update explicitly incorporates sequential inputs, allowing the model to process time-series data.

While the formulation in Equation \ref{rnn_equation} defines a vanilla RNN, training such networks through backpropagation through time is notoriously challenging for long sequences due to the issues of vanishing and exploding gradients \cite{graves2012long}. To address these optimization difficulties and enable the learning of long-term dependencies, researchers introduced gated RNN architectures. The most popular examples are the Long Short-Term Memory (LSTM) network \cite{hochreiter1997long} and the Gated Recurrent Unit (GRU) \cite{dey2017gate}. These models augment the basic recurrence with gating mechanisms that regulate the flow of information, allowing the network to decide which past information to keep or forget. By incorporating an internal cell state and gating signals, LSTMs (and similarly GRUs) can preserve gradients over long time lags, effectively mitigating the vanishing gradient problem and capturing longer-range temporal dependencies.

\subsection{Neural differential equations}\label{NCDEs}

\paragraph{Neural Ordinary Differential Equations (NODEs)}
Chen et al. \cite{chen2018neural}  first introduced NODEs, demonstrating that they can be viewed as the infinite-depth limit of residual networks (ResNets). A NODE models the evolution of a hidden state as the solution to an initial value problem. Specifically, the state $\boldsymbol{z}(t)$ evolves according to
\begin{equation}\label{node_equation}
\dot{\boldsymbol{z}}(t) = {\hat{f}}(t,\boldsymbol{z}(t);\boldsymbol{\theta}), \quad \boldsymbol{z}(t_0) = \boldsymbol{z}_0,
\end{equation}
where:
\begin{itemize}
    \item $\boldsymbol{z}(t) \in \mathbb{R}^d$ is the hidden state at time $t$,
    \item $\dot{\boldsymbol{z}}(t)$ denotes the time derivative of $\boldsymbol{z}(t)$,
    \item $\hat{f}(t,\boldsymbol{z}(t);\boldsymbol{\theta})$ is a vector field parameterized by $\boldsymbol{\theta}$ that governs the dynamics,
    \item $\boldsymbol{\theta}$ comprises the learnable parameters,
    \item $t_0$ is the initial time and $\boldsymbol{z}_0$ the corresponding initial state.
\end{itemize}
In fact, a ResNet’s layer-wise update $\boldsymbol{z}_{n+1}=\boldsymbol{z}_n+\boldsymbol{\hat{f}}(\boldsymbol{z}_{n};\boldsymbol{\theta)}$ resembles a forward Euler step of the ODE (see Equation \ref{node_equation}); letting the step size tend to zero leads to the continuous model. The NODE formulation thus forms a “continuous-depth” neural network that flows smoothly through hidden states, interacting seamlessly with the manifold hypothesis, as it describes a flow along which to evolve the data manifold \cite{kidger2020neural}. The expressivity of NODE has some limitations \cite{oh2025comprehensive}, because enforcing a Lipschitz vector field (to satisfy the Picard–Lindelöf conditions for existence and uniqueness) yields bijective, non‑intersecting flows that cannot represent mappings requiring trajectory crossings or other non‑diffeomorphic transformations.  Dupont et al. \cite{dupont2019augmented} addressed this by proposing Augmented NODEs, which add extra latent dimensions to $\boldsymbol{z}(t)$ so that the augmented system can represent more complex (non-diffeomorphic) transformations.

\paragraph{Neural Controlled Differential Equations (NCDEs)}
A limitation of NODEs is that they are initial value problems – once $\boldsymbol{z}(0)$ is set, the trajectory $\boldsymbol{z}(t)$ for $t>0$ is fixed by the ODE and cannot ingest new data points that arrive over time \cite{kidger2020neural}. In sequence modeling (time series), we often have streams of observations coming in; an RNN processes these by updating its hidden state at each observation. NODEs alone lack a mechanism to incorporate subsequent inputs after the initial time. NCDEs are built upon the theory of controlled differential equations, and can be viewed as the continuous-time analogue of RNNs. In other words, NCDEs extend the continuous framework of NODEs by incorporating external, time-dependent inputs into the evolution of the hidden state. The dynamics are given by
\begin{equation}\label{ncde_equation}
\dot{\boldsymbol{z}}(t) = \hat{f}(t,\boldsymbol{z}(t),\boldsymbol{X}(t);\boldsymbol{\theta})\,\dot{\boldsymbol{X}}(t), \quad \boldsymbol{z}(t_0) = \phi_{\text{init}}\left(\boldsymbol{X}(t_0)\right),
\end{equation}
where:
\begin{itemize}
    \item $\boldsymbol{z}(t) \in \mathbb{R}^d$ is the hidden state at time $t$,
    \item $\dot{\boldsymbol{z}}(t)$ is the time derivative of $\boldsymbol{z}(t)$,
    \item $\boldsymbol{X}(t)$ is the time-dependent input signal,
    \item $\dot{\boldsymbol{X}}(t)$ denotes the time derivative of $\boldsymbol{X}(t)$, serving as a control,
    \item $\hat{f}(t,\boldsymbol{z}(t),\boldsymbol{X}(t);\boldsymbol{\theta})$ is a parameterized vector field that modulates the impact of the input on the state dynamics,
    \item $\boldsymbol{\theta}$ includes the learnable parameters.
    \item $\phi_{\text{init}}$ is an initial function that maps the control at the initial time $t_0$ to the initial latent state $\boldsymbol{z}(t_0)$
\end{itemize}
If $\boldsymbol{X}(t)$ were constant, the NCDE reduces to a standard NODE; if $\hat{f}$ is zero, the hidden state remains fixed regardless of $\boldsymbol{X}$. Thus, NCDEs elegantly generalize NODEs by allowing continuous data injection.

Kidger et al. \cite{kidger2020neural} introduced Neural CDEs for modeling irregularly-sampled time series. In their setup, one first interpolates the discrete observations (e.g., with a piecewise linear or cubic spline) to obtain a continuous path $\boldsymbol{X}(t)$, then uses this $\boldsymbol{X}(t)$ as input to the NCDE. The NCDE’s hidden state $\boldsymbol{z}(t)$ is initialized at the start, a learned encoding of the first data point, and then evolves according to the following integral:
\begin{equation}\label{ncde_integral_form}
\boldsymbol{z}_t
  \;=\;
  \boldsymbol{z}_{t_0}
  \;+\;
  \int_{t_0}^{t} \hat{f}\!\bigl(\boldsymbol{z}_s;\boldsymbol{\theta}\bigr)\,\mathrm{d}\boldsymbol{X}_s
  \;=\;
  \boldsymbol{z}_{t_0}
  \;+\;
  \int_{t_0}^{t} \hat{f}\!\bigl(\boldsymbol{z}_s;\boldsymbol{\theta}\bigr)\,
             \frac{\mathrm{d}\boldsymbol{X}}{\mathrm{d}s}(s)\,\mathrm{d}s.
\end{equation}
Whenever an observation occurs, it influences $\boldsymbol{z}(t)$ via the increments $d\boldsymbol{X}$. Such a construction yields a continuous-time recurrent model — essentially a continuous analog of an RNN cell. A key advantage is that it naturally handles irregular intervals between observations: irregular gaps in time are handled by the ODE solver. Kidger et al. \cite{kidger2020neural} exhibited that NCDEs achieve state-of-the-art results on benchmarks, outperforming both pure NODEs (which can’t easily use new observations) and discrete RNN variants on irregular data. 

\subsection{Deep operator networks}\label{DeepONet}
The Deep Operator Network (DeepONet) is a neural architecture designed to learn mappings between infinite-dimensional function spaces, grounded in the universal approximation theorem for nonlinear operators \cite{lu2022comprehensive}. Its original formulation consists of two subnetworks: a \textit{branch network} that encodes the input function and a \textit{trunk network} that encodes the output coordinate. The operator $G: \mathcal{U} \to \mathcal{V}$ is represented as an inner-product coupling of these subnetworks. In particular, DeepONet models the output $G(u)(y)$ at any point $y$ as a linear combination of trunk outputs weighted by branch outputs, expressed as
\begin{equation}\label{node_equation}
G(u)(y) \approx \sum_{k=1}^p b_k\left(u(x_1, \dots, x_m)\right)\,t_k(y),
\end{equation}
where $b_k$ are the branch network outputs (depending on the input function $u$ sampled at points $x_i$), and $t_k$ are the trunk network outputs (depending on the query location $y$) \cite{kobayashi2024deep}. This construction directly follows the operator-valued universal approximation theory, which guarantees that such a network can approximate any continuous nonlinear operator to arbitrary accuracy, provided that the network has sufficient capacity. In other words, the trunk network in DeepONet can be interpreted as learning a set of nonlinear basis functions for the output space, while the branch network learns to map an input function to the appropriate coefficients for these basis functions\cite{sharma2024ensemble}. This separable structure — where one network handles function inputs and another handles output coordinates — is the key to DeepONet's expressive power and generalization in operator learning.

\section{NCDE-DeepONet}\label{ncdedeeponet}

\subsection{Overall setup}\label{ncdedeeponet_setup}
\begin{figure}[htbp]
  \centering
  \includegraphics[width=1.0\textwidth]{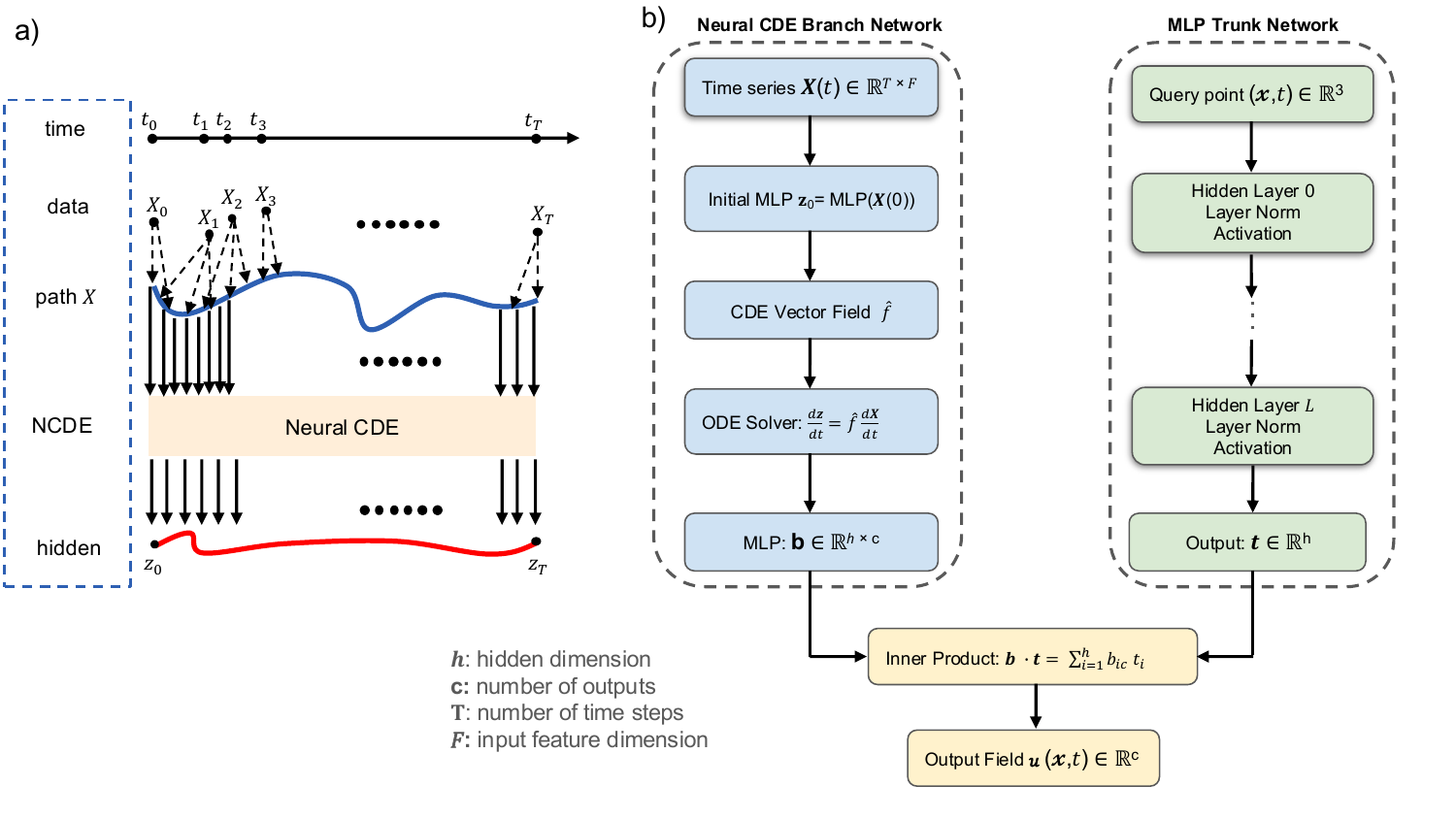}
  \caption{(a) Neural Controlled Differential Equation (NCDE) processing of irregular time series. Some data process is observed at times $t_0, t_1, \ldots, t_T$ to give observations $X_0, X_1, \ldots, X_T$. It is otherwise unobserved. In contrast to discrete models, the hidden state of the Neural CDE model has continuous dependence on the observed data through the interpolated path $X$. (b) NCDE-DeepONet architecture: the branch network maps irregular time series to embeddings via a Neural CDE; the trunk network embeds spatio-temporal query points $(x,t)$; their inner product yields the predicted field $u(x,t) \in \mathbb{R}^c$.}
  \label{fig_ncde_architecture}
\end{figure}

We propose a novel operator learning framework, termed NCDE-DeepONet, which integrates a Neural Controlled Differential Equation (NCDE) into the branch network of the standard DeepONet architecture, while the trunk network receives both spatial coordinates and time as input (see Figure \ref{fig_ncde_architecture}). This design introduces two key departures from existing architectures, such as Sequential DeepONet (S-DeepONet) \cite{he2024predictions}, which employs a gated recurrent unit (GRU) in the branch and omits temporal information from the trunk. First, by adopting an NCDE-based branch, our model is inherently equipped to process irregularly sampled input signals, which are common in real-world dynamical systems, by treating them as continuous control paths; thus, it offers a principled mechanism for temporal encoding. Second, by incorporating time into the trunk input, the model enables direct spatiotemporal querying of the learned operator, allowing the solution to be evaluated at arbitrary locations and time instances. This decoupling between the input signal and spatiotemporal query points enhances the model's ability to generalize better across space and time, facilitating downstream tasks such as interpolation and forecasting in continuous domains. These design choices endow NCDE-DeepONet with enhanced representational capacity for complex temporal systems, particularly those characterized by irregular input patterns and continuous-time evolution.

Building on the conceptual structure illustrated in Figure \ref{fig_ncde_architecture}, the NCDE-DeepONet architecture comprises two decoupled yet jointly trained components: an NCDE-based branch network and an MLP-based trunk network. The branch is designed to encode multivariate time-series signals $\boldsymbol{X}(t)$ into a compact latent representation through a sequence of operations grounded in continuous-time modeling. Formally, an NCDE can be regarded as the continuous-time limit of a recurrent neural network; discretising the integral in Equation \ref{ncde_equation} with an explicit Euler step recovers an ordinary RNN cell \cite{choi2022graph}. First, the raw input trajectory is interpolated to a continuous path using cubic Hermite interpolation, depending on the configuration. This interpolation defines the control signal $\boldsymbol{X}(t)$ (see Figure \ref{fig_ncde_architecture}a). Because discrete observations are first transformed into a continuous control path, the branch naturally accommodates missing or unevenly spaced samples—the interpolation step absorbs such gaps without any architectural modifications. This contrasts sharply with discrete recurrent architectures, such as GRUs, which require special handling for irregular sampling, including imputation or time-aware gating mechanisms. An initial function $\phi_{\text{init}}$ maps the control at the initial time $t_0$ to the initial latent state $\boldsymbol{z}(t_0)$, which is then evolved forward in time using a learned vector field $f_\theta$ within the NCDE framework (see Equation \ref{ncde_equation}). The hidden state $\boldsymbol{z}(T)$ at the terminal time $t=T$ is computed using a numerical ODE solver (e.g., Tsit5 or Dopri8) with adaptive error control and step size. This terminal latent state is subsequently passed through a linear projection network to yield the branch output vector $\boldsymbol{b} \in \mathbb{R}^h$, which encodes the entire observed input trajectory as a fixed-length representation.

\begin{figure}[htbp]
  \centering
  \includegraphics[width=1.0\textwidth]{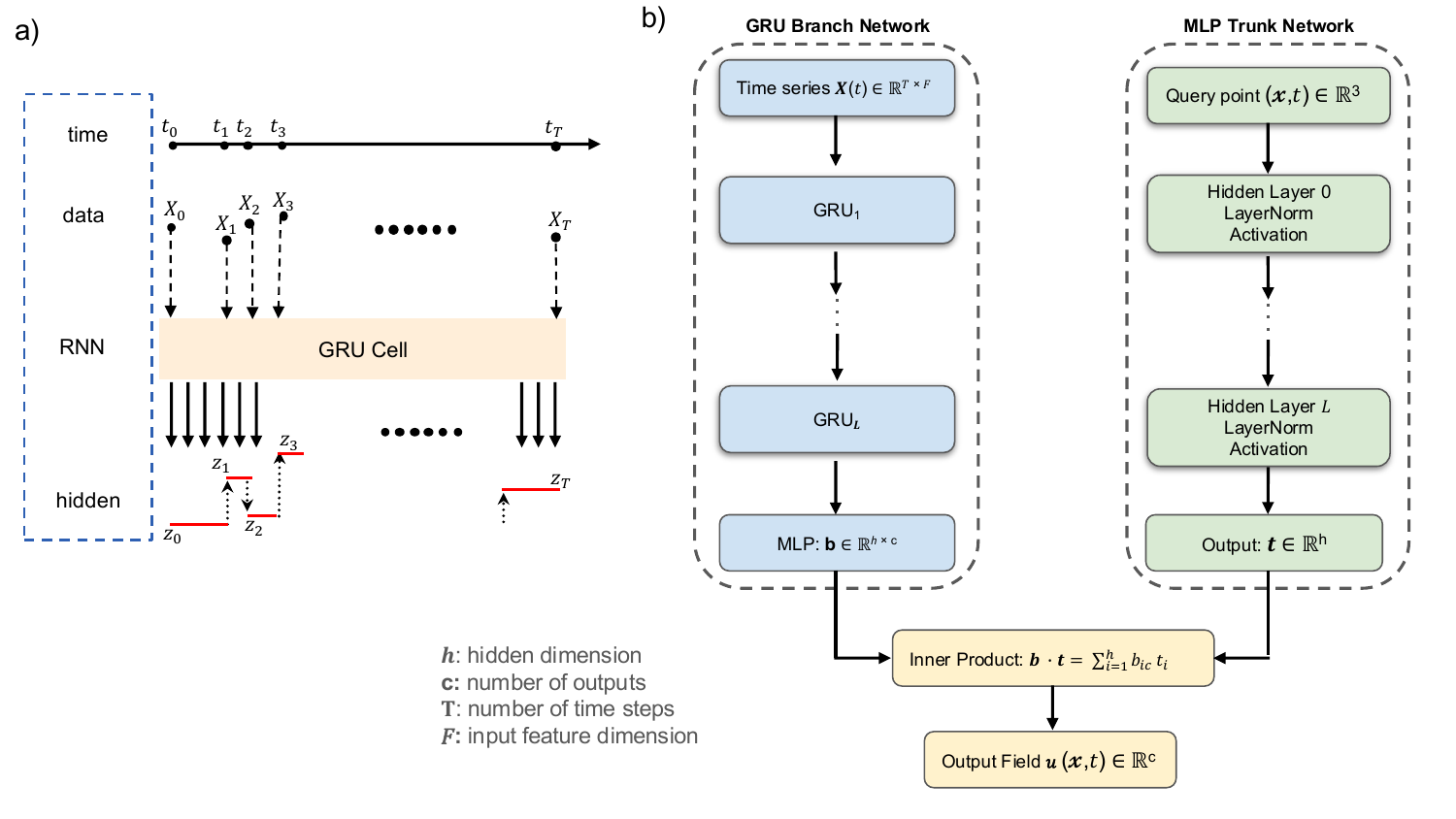}
  \caption{(a) Gated Recurrent Unit (GRU) processing of irregular time series. Some data process is observed at times $t_0, t_1, \ldots, t_T$ to give observations $X_0, X_1, \ldots, X_T$. It is otherwise unobserved. The GRU modifies the hidden state at each observation with no evolution between observations (indicated by horizontal red lines). (b) GRU-DeepONet architecture: the branch network maps irregular time series to embeddings via a stacked GRU; the trunk network embeds spatio-temporal query points $(x,t)$; their inner product yields the predicted field $u(x,t) \in \mathbb{R}^c$.}
  \label{fig_gru_architecture}
\end{figure}

In parallel, the trunk network processes spatiotemporal coordinates $(\boldsymbol{x}, t)$ by embedding them into the same latent space $\mathbb{R}^h$ through a series of affine transformations and non-linearities with intermediate layer normalization. This yields a location- and time-dependent representation $\boldsymbol{t} \in \mathbb{R}^h$. The final output is computed as the inner product $\widehat{u}(\boldsymbol{x}, t) = \langle \boldsymbol{b}, \boldsymbol{t} \rangle$, effectively coupling the temporally encoded trajectory with the spatiotemporal query. This decoupling allows the model to generalize across different spatiotemporal configurations, including unseen times and sensor locations, without retraining or re-encoding the input signal. The model is trained using standard supervised loss over sampled mini-batches of input trajectories and spatiotemporal query–target pairs. During training and evaluation, spatiotemporal query points $(\boldsymbol{x}, t)$ are randomly sampled from a discrete set defined by the underlying FEM-generated dataset. While this set is finite and fixed a priori, the sampling is not restricted to a structured grid, allowing for dense and stochastic domain coverage. Crucially, the trained model supports prediction at arbitrary locations and time instances post-training because the trunk network receives arbitrary-valued $(\boldsymbol{x}, t)$ as input, while the branch network can handle irregular input signals. In addition to the NCDE-DeepONet, we develop a DeepONet with GRU layers in the branch instead of the NCDE for comparison purposes. Fig. \ref{fig_gru_architecture} illustrates the GRU-DeepONet. Unlike the NCDE-DeepONet, which evolves its hidden state continuously through controlled differential equations, the GRU-DeepONet processes data through discrete recurrent updates at each observation time. This fundamental difference means that the GRU-DeepONet treats all observations as equally spaced in time, ignoring the actual temporal gaps between measurements, whereas the NCDE approach naturally incorporates these irregular time intervals through its continuous formulation. However, since both DeepOnets have $x$ and $t$ in the trunk, both support prediction at arbitrary locations and time instances post-training.

\subsection{Training considerations}\label{ncdedeeponet_training}
As discussed in Section \ref{ncdedeeponet_setup}, the branch of the NCDE-DeepONet has an NCDE network. Training NODEs and NCDEs requires solving ODEs numerically as part of the forward and backward passes. A variety of integration schemes can be used, from basic fixed-step methods to sophisticated adaptive solvers. Explicit fixed-step methods, such as the forward Euler method or the classical Runge–Kutta (RK4) method, offer straightforward implementations but may require small step sizes for accuracy or stability, particularly for systems with stiff dynamics. In practice, higher-order adaptive solvers are commonly employed to adjust step sizes and automatically control local truncation error. Adaptive step-size integrators aim to meet specified error tolerances by rejecting or reducing steps that would introduce excessive error \cite{kidger2021hey}. This ensures a specified level of accuracy while using as few function evaluations as necessary. However, adaptivity means that the number of ODE function evaluations (and thus the computational cost) can vary and may increase significantly if the learned dynamics are complex or stiff. In the context of NCDEs, an additional step is to construct a continuous interpolation of the input path (often a natural cubic spline through the discrete observations) to serve as the driving signal \cite{kidger2020neural}. Once this interpolation is in place, the NCDE can be solved using the same ODE solvers as those for NODEs. When the vector field $\hat{f}$ is Lipschitz-continuous, the NCDE admits a unique solution that depends continuously on the input path, a property that also guarantees stable gradient propagation during training \cite{choi2022graph}.

Training a Neural Controlled Differential Equation (NCDE) requires backpropagating through the continuous hidden-state trajectory, which can be memory-intensive if every intermediate state is stored. This differs from training discrete RNN architectures, such as GRUs, where gradients are propagated through a finite sequence of hidden states rather than a continuous trajectory. In this work, we utilize the adjoint sensitivity method to compute gradients for NCDEs efficiently, mirroring the approach employed in Neural ODEs while accounting for the input control signal. The key idea of the adjoint method is to augment this system with additional variables that track sensitivities, and then integrate this augmented system backward in time to obtain gradients, avoiding the need to store intermediate $\boldsymbol{z}(t)$ values in memory \cite{chen2018neural}. In a standard Neural ODE (without an external control), the continuous adjoint method uses an adjoint state to propagate gradients backward in time through the ODE solver. By contrast, in an NCDE, the dynamics are driven by a time-varying control path $X(t)$, which modifies the adjoint equations while preserving the core idea of backward-in-time gradient integration \cite{kidger2020neural}. Specifically, let $\boldsymbol{z}(t)$ be the NCDE hidden state satisfying a forward CDE (see Equation \ref{ncde_equation}). To compute gradients of a loss $\mathcal{L}$ (e.g., a mean squared error) with respect to parameters, we define an adjoint process $\boldsymbol{a}(t) = \frac{\partial \mathcal{L}}{\partial \boldsymbol{z}(t)}$, which represents the instantaneous sensitivity of the loss to the state at time $t$. The adjoint $\boldsymbol{a}(t)$ evolves backwards in time according to the backward CDE (adjoint equation) \cite{kidger2022neural}:
\begin{equation}\label{adjoint_cde}
\frac{d\boldsymbol{a}(t)}{dt} \;=\; 
-\,\Bigl(\tfrac{\partial \hat{f}}{\partial \boldsymbol{z}}\bigl(\boldsymbol{z}(t)\bigr)\Bigr)^{\!\top}
\,\boldsymbol{a}(t)\,\dot{\boldsymbol{X}}(t),
\end{equation}
with terminal condition $\boldsymbol{a}(T) = \frac{\partial \mathcal{L}}{\partial \boldsymbol{z}(T)}$. Here, $(\partial_{\boldsymbol{z}}\hat{f}( \boldsymbol{z}(t) ))^T \boldsymbol{a}(t)$ denotes the transpose of the Jacobian of $\hat{f}$ (with respect to $\boldsymbol{z}$), multiplied by the adjoint vector $\boldsymbol{a}(t)$. $\dot{\boldsymbol{X}}(t)$ scales this influence by the rate of change of the input. This equation accumulates the gradient backpropagation signal over time: as we integrate from $t=T$ down to $t=0$, the adjoint $\boldsymbol{a}(t)$ tracks how perturbations in the hidden state at time $t$ would affect the final loss \cite{kidger2021hey}. The negative sign ensures that $\boldsymbol{a}(t)$ evolves along the direction of steepest descent for the loss during the backward integration. This construction is analogous to the Neural ODE adjoint ($\dot{\boldsymbol{a}} = -\boldsymbol{a}^T \partial \hat{f}/\partial \boldsymbol{z}$), but with the important distinction that the input path derivative $\dot{\boldsymbol{X}}(t)$ appears as a factor, reflecting the controlled nature of the system.

During the same backward sweep, we introduce a parameter-gradient accumulator $\boldsymbol{p}(t)$ (representing $\partial \mathcal{L}/\partial \boldsymbol{\theta}$ accumulated from time $T$ down to $t$) \cite{matsubara2021symplectic}. Writing the stacked vector
$\mathbf{y}(t)=\bigl[\boldsymbol{z}(t), \boldsymbol{a}(t), \boldsymbol{p}(t)\bigr]^{T}$, the full augmented adjoint system reads:
\begin{equation}\label{augmented_ncde}
\frac{d\mathbf{y}(t)}{dt} =
\begin{bmatrix}
\hat{f}\bigl(\boldsymbol{z}(t)\bigr)\,\dot{\boldsymbol{X}}(t) \\[6pt]
-\bigl(\partial_{\boldsymbol{z}}\hat{f}\bigl(\boldsymbol{z}(t)\bigr)\bigr)^{\!\top}\boldsymbol{a}(t)\,\dot{\boldsymbol{X}}(t) \\[8pt]
-\boldsymbol{a}(t)^{\!\top}\,\partial_{\boldsymbol{\theta}}\hat{f}\bigl(\boldsymbol{z}(t)\bigr)\,\dot{\boldsymbol{X}}(t)
\end{bmatrix},
\quad
\mathbf{y}(T) =
\begin{bmatrix}
\boldsymbol{z}(T) \\[4pt]
\displaystyle \frac{\partial \mathcal{L}}{\partial \boldsymbol{z}(T)} \\[4pt]
\mathbf{0}
\end{bmatrix}.
\end{equation}

Using the chain rule exactly as in the NODE derivation of \cite{chen2018neural}, one obtains the NCDE counterpart
\begin{equation}\label{eq:ncde_intgrad}
\nabla_{\!\boldsymbol{\theta}}\mathcal{L}
\;=\;
\int_{t_0}^{T}
\boldsymbol{a}(t)^{\top}\,
\partial_{\boldsymbol{\theta}}\hat{f}\!\bigl(\boldsymbol{z}(t)\bigr)\,
\dot{\boldsymbol{X}}(t)
\,\mathrm{d}t,
\end{equation}
where the adjoint $\boldsymbol{a}(t)$ satisfies the backward CDE Equation \ref{adjoint_cde}.
Equation \ref{eq:ncde_intgrad} is recovered automatically by the augmented system (see Equation \ref{augmented_ncde}). Since $\tfrac{\mathrm{d}}{\mathrm{d}t}\,\boldsymbol{p}(t) = -\,\boldsymbol{a}(t)^{\top}\partial_{\boldsymbol{\theta}}\hat{f}\bigl(\boldsymbol{z}(t)\bigr)\dot{\boldsymbol{X}}(t)$,
integrating backward from $t=T$ to $t_0$ gives
$\boldsymbol{p}(t_0) = \nabla_{\!\boldsymbol{\theta}}\mathcal{L}$,
in agreement with Equation \ref{eq:ncde_intgrad}.

Integrating Eq.\eqref{augmented_ncde} backward from $t=T$ to $t_0$ carries out three tasks in a single sweep \cite{matsubara2021symplectic}:
\begin{itemize}
    \item \textbf{Trajectory reconstruction:} the solver re-integrates the forward state $\boldsymbol{z}(t)$ on demand, so no full trajectory is stored.
    \item \textbf{Adjoint propagation:} the loss sensitivity $\boldsymbol{a}(t)$ is transported backward through time.
    \item \textbf{Gradient accumulation:} the parameter accumulator $\boldsymbol{p}(t)$ integrates $\nabla_{\!\boldsymbol{\theta}}\mathcal{L}$ throughout the interval.
\end{itemize}
Once the backward integration reaches $t_0$, we recover
$\boldsymbol{p}(t_0)=\nabla_{\!\boldsymbol{\theta}}\mathcal{L}$—exactly as shown in Equation \ref{eq:ncde_intgrad} \cite{matsubara2021symplectic}. This gradient can be passed directly to any gradient-based optimizer. The trade-off for this constant-memory strategy is a second ODE solve (the backward pass) plus a few Jacobian–vector products, a computational overhead that is typically outweighed by the substantial memory savings and the ability to train on long or finely resolved time series \cite{zhuang2020adaptive}.

Unlike the NCDE branch, the trunk network is implemented as a finite stack of affine transformations, element-wise nonlinearities, and layer normalization blocks, all of which are amenable to standard reverse-mode differentiation. During back-propagation, the gradient at each affine layer is mapped by the corresponding weight matrix, whereas the gradient at each activation function is scaled component-wise by the derivative of that function evaluated at the pre-activation value. Because each nonlinearity operates independently on its input, the associated Jacobian is diagonal, indicating that every output neuron depends exclusively on its own pre-activation. Layer normalization is likewise fully differentiable: its per-sample mean–variance normalization followed by learnable gain and bias parameters transmits gradients across all features and to the gain/bias variables \cite{ba2016layer}. Finally, the model’s scalar prediction is obtained as an inner product, $\boldsymbol{u}= \langle \boldsymbol{b}, \boldsymbol{t}\rangle$, so the loss gradient decomposes cleanly into complementary terms for the branch and trunk, enabling their parameters to be updated synchronously within the same optimization step.
\section{Numerical examples}\label{results}

In this section, we demonstrate the effectiveness of the proposed NCDE-DeepONet architecture through numerical experiments on three problems governed by partial differential equations: the transient Poisson equation, elastodynamics, and transient thermoelasticity. The corresponding subsections provide detailed descriptions of each problem setting. All examples employ identical hyperparameters across problems, demonstrating the framework's robustness without problem-specific tuning (see Appendix A.1 for details).

For all examples, we evaluate the model performance using the relative $L_2$ error defined as:
\begin{equation}
\mathcal{E}_{\text{rel}} = \frac{\|\mathbf{y}_{\text{true}} - \mathbf{y}_{\text{pred}}\|_2}{\|\mathbf{y}_{\text{true}}\|_2}
\label{eq:relative_error}
\end{equation}
where $\mathbf{y}_{\text{true}}$ and $\mathbf{y}_{\text{pred}}$ denote the true and predicted field values across all spatial locations and time steps, and $\|\cdot\|_2$ represents the $L_2$ norm.

The following subsections present three numerical examples to comprehensively evaluate the NCDE-DeepONet framework. The transient Poisson problem (Section \ref{poisson_example}) serves as a foundational test case for scalar field prediction, where we compare two architectural variants—spatiotemporal and spatial-only trunk networks—to assess the benefits of incorporating temporal information directly in the trunk. The elastodynamics problem (Section \ref{elastodynamics}) extends the evaluation to vector fields with two displacement components, providing a direct comparison between our NCDE-based approach and a GRU-DeepONet baseline to demonstrate the advantages of continuous-time neural differential equations. Finally, the thermoelasticity problem (Section \ref{thermoelasticity}) challenges the framework with a coupled multiphysics system involving both mechanical displacements and temperature fields, while showcasing the NCDE's unique capability to handle variable input sampling rates without retraining. Together, these examples progressively demonstrate the framework's versatility, accuracy, and practical advantages for learning solution operators of time-dependent PDEs.

\subsection{Poisson}\label{poisson_example}

\begin{figure}[!htb]
    \centering
    \includegraphics[width=0.6\textwidth]{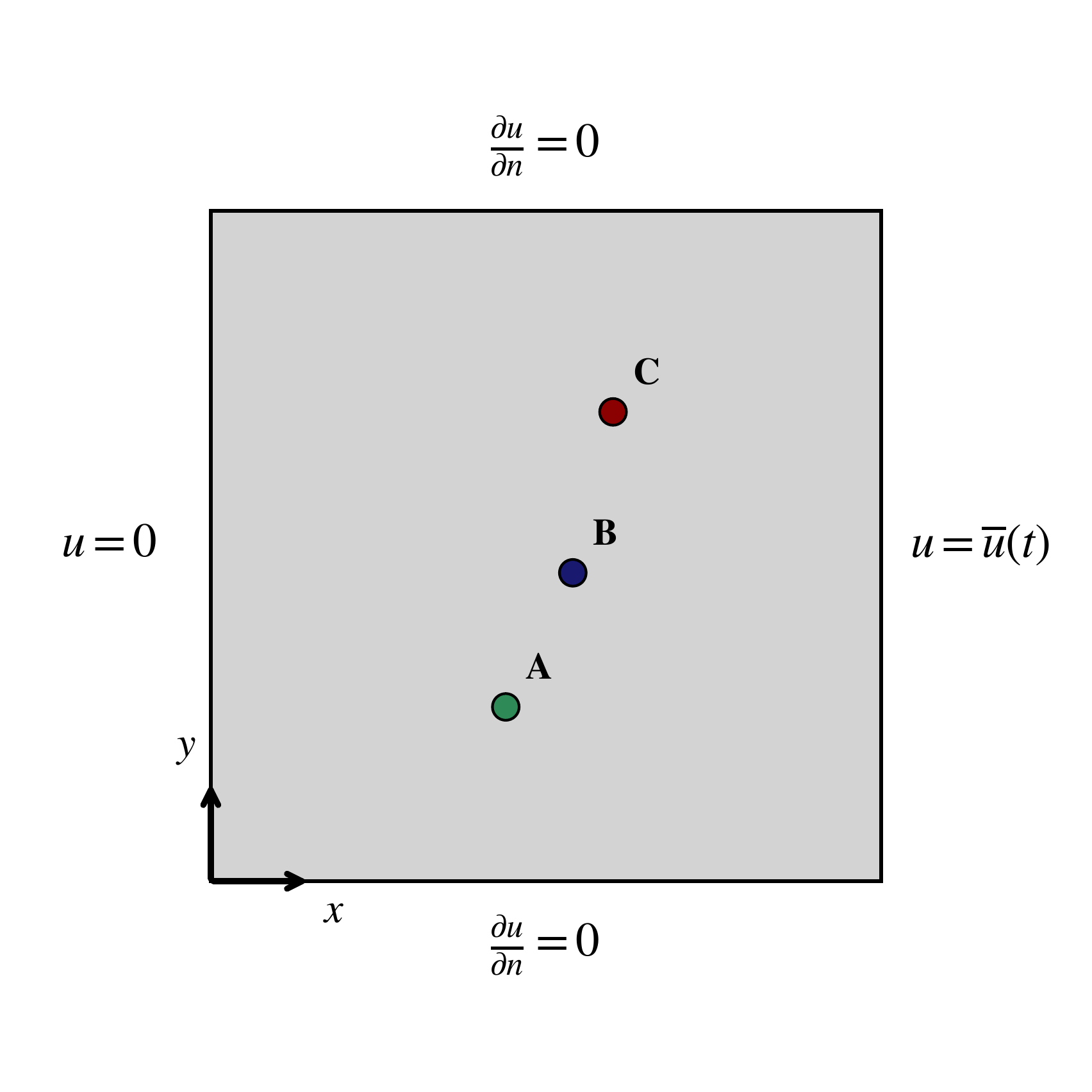}
    \caption{Schematic of the geometry and boundary conditions of the Poisson example.}
    \label{Figure_Schematic_poisson}
\end{figure}

Let $\Omega$ be a homogeneous unit square.
A time-dependent Dirichlet condition $u=\overline{u}(t)$ is prescribed on the right edge, while the left edge is fixed to zero.
The top and bottom edges are subject to homogeneous Neumann (flux-free) conditions.
Fig. \ref{Figure_Schematic_poisson} sketches the domain and these boundary conditions.
The scalar field $u(\mathbf{x},t)$ is governed by the transient Poisson equation:
\begin{equation}\label{poisson_equation}
\begin{aligned}
\frac{\partial u}{\partial t} - \Delta u &= f  \quad \; \; \; \; \! \quad \boldsymbol{x}\in\Omega \text{, } t\in (0, T],\\
u(x, 0) &= u_0(x) \quad  \; \boldsymbol{x}\in\Omega\\
u &= \overline{u}\left(t\right) \; \; \; \quad \! \boldsymbol{x}\in\Gamma_{u}\\
\nabla u \cdot \mathbf{n} &= 0 \quad \; \; \; \quad \; \boldsymbol{x} \in \Gamma_t
\end{aligned}
\end{equation}
where $\nabla^{2}$ is the Laplacian and $f=0$ in this example. $u_0(x)$ is the initial condition, which is set to zero, i.e., $u_0(x) = 0$.

\paragraph{Data generation and training}
We generated a dataset of 3,750 samples by varying the time-dependent Dirichlet boundary condition $\overline{u}(t)$ applied at the right edge. Each sample consists of the input signal $\overline{u}(t)$ and the corresponding scalar field evolution $u(\mathbf{x},t)$ over the time interval, computed using the finite element method with evenly spaced time increments. The dataset was split into 3,375 samples for training and 375 samples for testing. The NCDE-DeepONet was trained to learn the mapping from the time-varying boundary condition to the complete spatiotemporal evolution of the scalar field $u$.

\begin{figure}[!htb]
    \centering
    \includegraphics[width=1.0\textwidth]{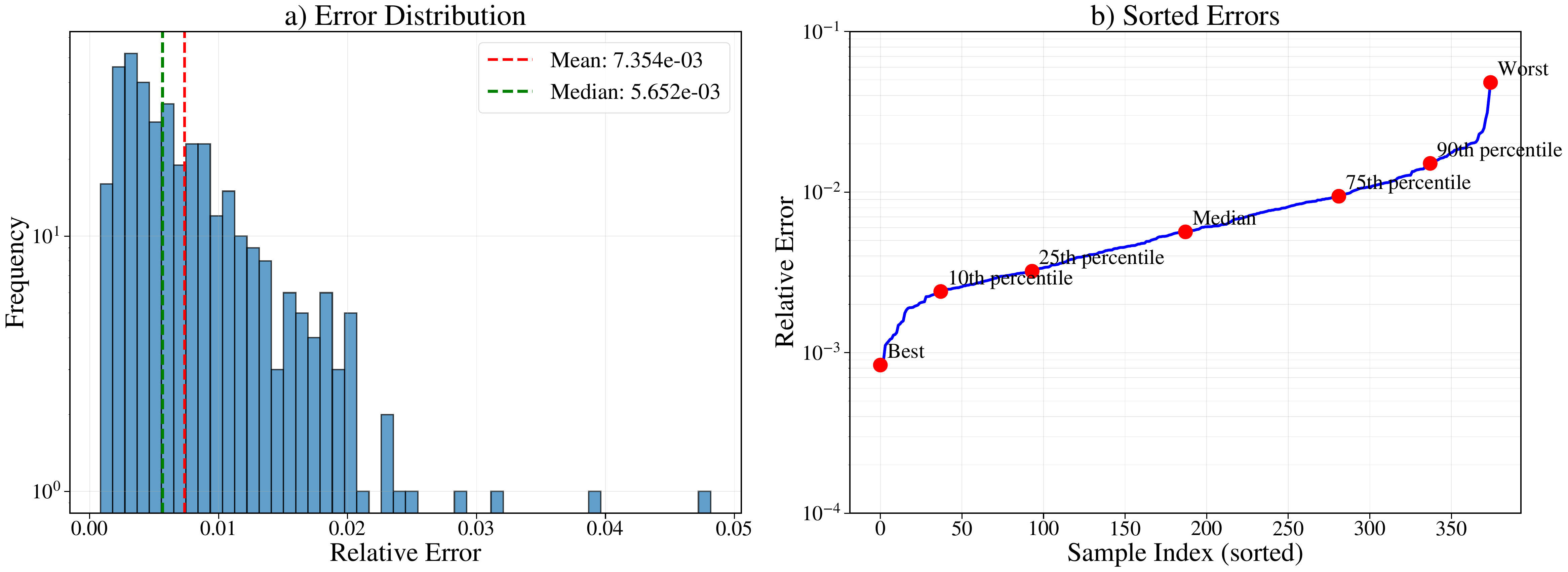}
    \caption{Error analysis of the NCDE-DeepONet model on the Poisson test dataset. (a) Distribution of relative errors showing a right-skewed profile with mean and median values indicated. (b) Sorted relative errors across all test samples with key percentiles marked, demonstrating consistent model performance.}
    \label{fig:poisson_error}
\end{figure}

\paragraph{Error analysis}
Fig. \ref{fig:poisson_error} presents a comprehensive error analysis of the trained model on the test dataset using the relative error metric defined in Eq. \eqref{eq:relative_error}. The error distribution (Fig. \ref{fig:poisson_error}a) exhibits a right-skewed profile with a mean relative error of $7.35 \times 10^{-3}$ and a median of $5.65 \times 10^{-3}$. The concentration of errors at lower values indicates consistent model performance across the majority of test cases. The sorted error plot (Fig. \ref{fig:poisson_error}b) reveals that the model achieves high accuracy for most samples, with relative errors below $10^{-2}$ for approximately 80\% of the test cases. The worst-case error remains bounded below $5 \times 10^{-2}$, demonstrating robust generalization without significant outliers.

\begin{figure}[!htb]
    \centering
    \includegraphics[width=1.0\textwidth]{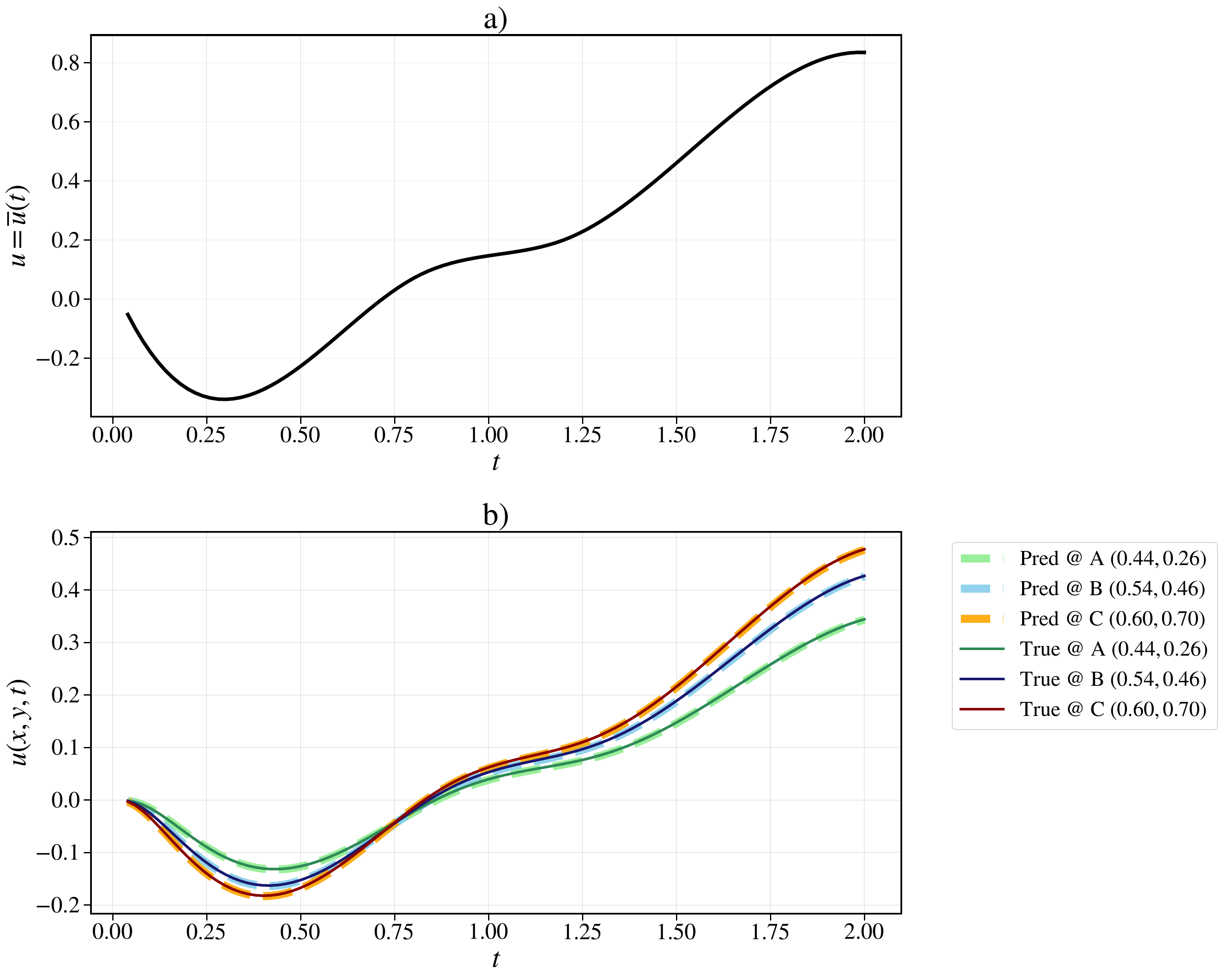}
    \caption{Time evolution of the Poisson field for the median error test case. True (solid lines) and predicted (dashed lines) field values at three spatial locations.}
    \label{fig:poisson_time_evolution}
\end{figure}

\paragraph{Field predictions} To illustrate the model's predictive capabilities, we analyze the performance across different error percentiles.  The corresponding input signals $\overline{u}(t)$ for various error percentiles, showcasing diverse temporal characteristics, are provided in Fig. \ref{fig:poisson_input_signals} in Appendix A.2.
Fig. \ref{fig:poisson_heatmap_median} shows the spatial field distribution for the median error case at $t = 2$s. The NCDE-DeepONet accurately captures the solution, with the predicted field closely matching the reference. Absolute errors remain on the order of $10^{-3}$—two orders of magnitude smaller than the field values. The most considerable discrepancies occur near the right boundary where the time-varying Dirichlet condition is imposed, as expected due to the dynamic nature of this boundary. Additional visualizations for other error percentiles, including the best, worst, and intermediate cases, are presented in Appendix A.2. To further demonstrate the model's temporal accuracy, Fig. \ref{fig:poisson_time_evolution} presents the time evolution of the field for the median error case. This figure displays the field evolution at three spatial locations within the domain. The NCDE-DeepONet predictions (dashed lines) closely track the reference solutions (solid lines) throughout the entire time interval, accurately capturing both the transient response to the changing boundary condition and the spatial variation in the field magnitude.

\begin{figure}[!htb]
    \centering
    \includegraphics[width=1.0\textwidth]{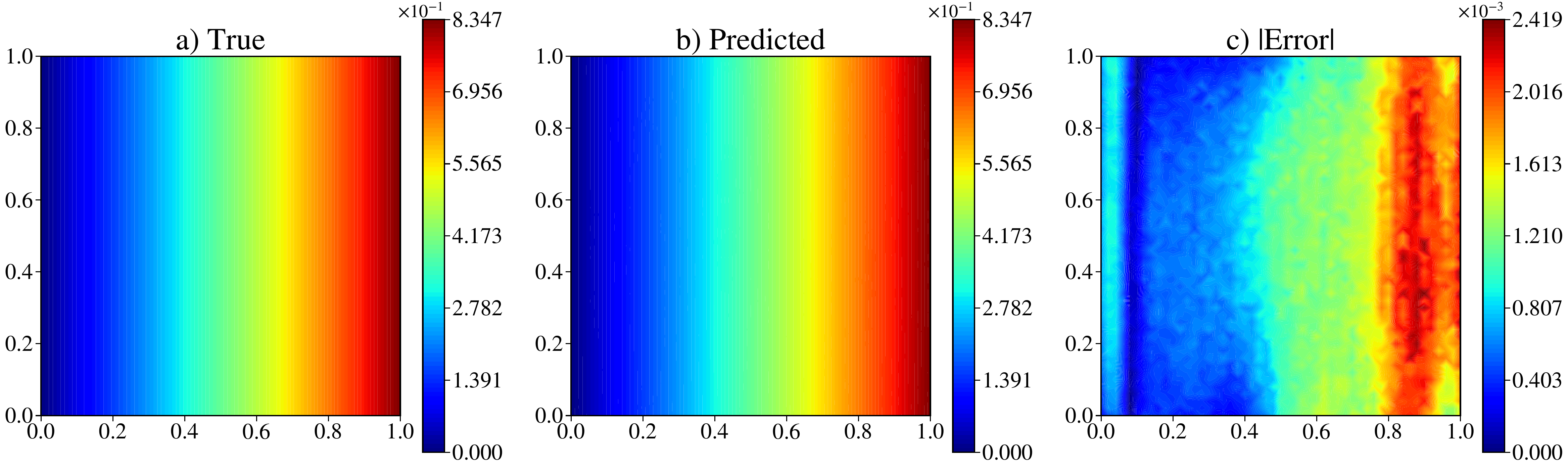}
    \caption{Poisson field predictions at $t = 2$s for the median error test case. Columns show true field, NCDE-DeepONet prediction, and absolute error for the scalar field $u(\mathbf{x},t)$.}
    \label{fig:poisson_heatmap_median}
\end{figure}

To investigate the role of temporal information in the trunk network, we compare two architectural variants of NCDE-DeepONet:
\begin{enumerate}
    \item \textbf{Spatiotemporal trunk} (main architecture): The trunk network processes spatial coordinates $(x, y)$ along with time $t$, taking 3-dimensional inputs $(x, y, t)$.
    
    \item \textbf{Spatial-only trunk}: The trunk network processes only spatial coordinates $(x, y)$, while temporal dynamics are captured entirely by the branch network through the Neural CDE.
\end{enumerate}
In the spatial-only variant, the branch network outputs time-evolving features $\mathbf{b} \in \mathbb{R}^{T \times h}$ for $T$ time steps, while the trunk produces spatial basis functions $\mathbf{t}(x,y) \in \mathbb{R}^{h}$. The final prediction is computed as:
\begin{equation}
    u(x, y, t_j) = \sum_{i=1}^{h} b_{ji} \cdot t_i(x, y)
\end{equation}
where $j \in \{1, 2, ..., T\}$ indexes the discrete time steps and $i$ indexes the hidden dimensions.

\begin{figure}[!htb]
    \centering
    \includegraphics[width=1.0\textwidth]{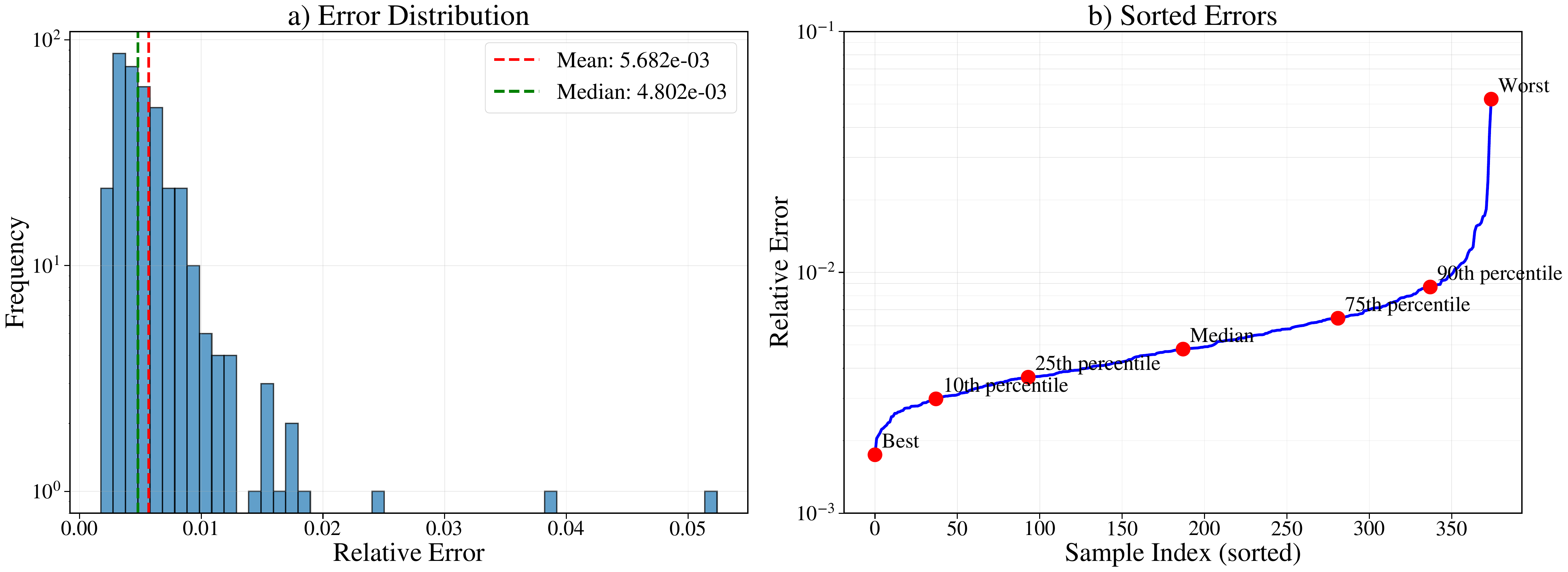}
    \caption{Error analysis of the spatial-only trunk NCDE-DeepONet model on the Poisson test dataset. (a) Distribution of relative errors showing a right-skewed profile with mean and median values indicated. (b) Sorted relative errors across all test samples with key percentiles marked.}
    \label{fig:poisson_error_spatial}
\end{figure}

Comparing the error analyses between the spatiotemporal trunk (Fig. \ref{fig:poisson_error}) and the spatial-only trunk (Fig. \ref{fig:poisson_error_spatial}) architectures reveals interesting performance characteristics. The spatial-only trunk architecture achieves slightly lower errors with a mean relative error of $5.68 \times 10^{-3}$ and median of $4.80 \times 10^{-3}$, compared to the spatiotemporal trunk's mean of $7.35 \times 10^{-3}$ and median of $5.65 \times 10^{-3}$. Both architectures exhibit similar error distribution profiles—right-skewed with the majority of samples achieving errors below $10^{-2}$, and worst-case errors remaining bounded below $5 \times 10^{-2}$. This demonstrates that both architectural choices provide robust and reliable predictions across the test dataset.

While the spatial-only trunk demonstrates competitive accuracy, the spatiotemporal trunk architecture offers significant advantages in practical applications:

\begin{enumerate}
    \item \textbf{Continuous temporal predictions}: The spatiotemporal trunk can evaluate the solution at arbitrary time points $t \in [0, T]$ through the continuous trunk input $(x, y, t)$. This enables high-resolution temporal reconstructions and predictions at any desired time instant without retraining.
    
    \item \textbf{Fixed temporal resolution limitation}: The spatial-only trunk is constrained to predict only at the discrete time points $t_j$ used during training. The branch network outputs are fixed at $T$ time steps, limiting temporal resolution to the training discretization. Obtaining predictions at intermediate times would require interpolating the branch features $b_{ji}$ or retraining the model with a finer temporal discretization.
    
\end{enumerate}

Therefore, despite the marginal accuracy advantage of the spatial-only variant on this specific problem, we adopt the spatiotemporal trunk as our primary architecture due to its superior versatility and practical applicability in real-world scenarios where temporal flexibility is crucial.

\subsection{Elastodynamics}\label{elastodynamics}

\begin{figure}[!htb]
    \centering
    \includegraphics[width=0.5\textwidth]{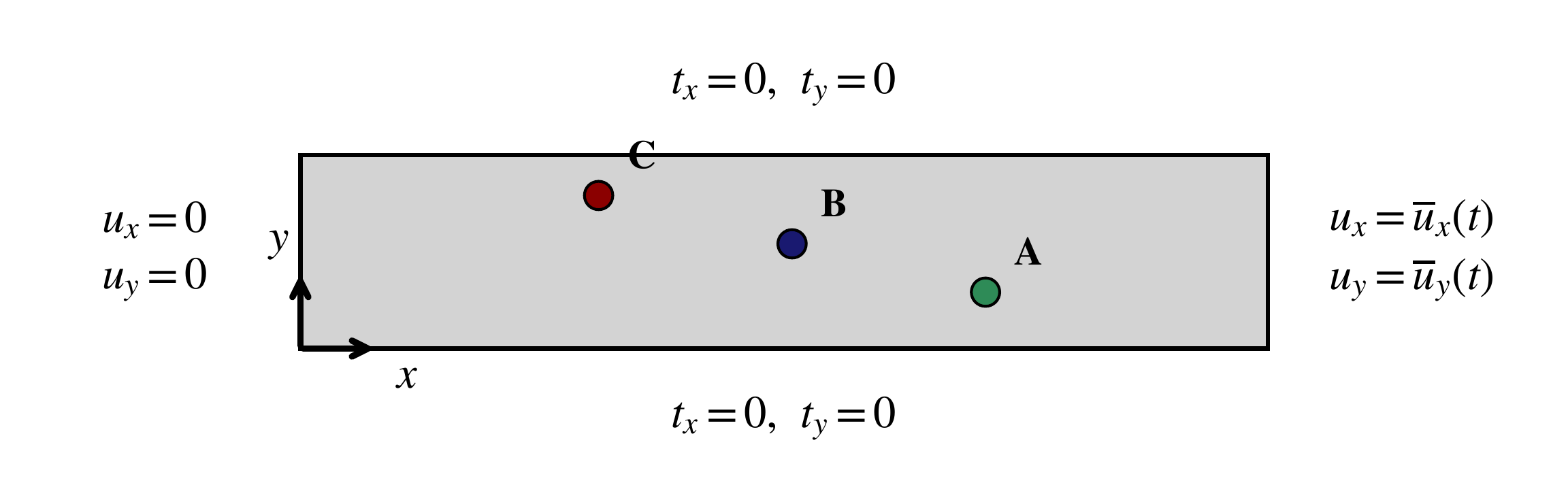}
    \caption{Schematic of the geometry and boundary conditions of the elastodynamics example.}
    \label{Figure_Schematic_elasto}
\end{figure}

A homogeneous rectangular plate $\Omega\subset\mathbb{R}^2$ of width
$L_x=1$ and height $L_y=0.2$ is considered under plane-stress elastodynamics. The left edge is fully fixed, enforcing $u_x=u_y=0$, while the right edge follows a prescribed time-dependent displacement
$\bar{\mathbf u}(t)=\bigl(\bar u_x(t),\bar u_y(t)\bigr)^{\!\top}$. The top and bottom edges are traction-free. Fig. \ref{Figure_Schematic_elasto} illustrates these boundary conditions. The displacement field
$\mathbf u(\mathbf x,t)=(u_x,u_y)^{\!\top}$ obeys
\begin{equation}\label{eq:elasto_dyn_ps}
\begin{aligned}
\rho\,\ddot{\boldsymbol{u}}-\nabla\!\cdot\!\boldsymbol{\sigma} &= \mathbf 0 &&\text{in } \Omega,\; t\in(0,T],\\
\boldsymbol{u}(\mathbf x,0) &= \mathbf 0,\quad
\dot{\boldsymbol{u}}(\mathbf x,0)=\mathbf 0 &&\text{in } \Omega,\\
\boldsymbol{u} &= \bar{\boldsymbol{u}}(t) &&\text{on } \Gamma_{u},\\
\boldsymbol{\sigma}\,\mathbf n &= \mathbf 0 &&\text{on } \Gamma_{t},
\end{aligned}
\end{equation}
with plane-stress constitutive law
\begin{equation}\label{elasto_const}
\begin{aligned}
\boldsymbol{\sigma}
  = \lambda\,\operatorname{tr}\boldsymbol{\varepsilon}\,\mathbf I
    + 2\mu\,\boldsymbol{\varepsilon},\qquad
\lambda=\frac{E\nu}{(1+\nu)(1-\nu)},\;\;
\mu=\frac{E}{2(1+\nu)},
\end{aligned}
\end{equation}
where
\(\boldsymbol{\varepsilon}=\tfrac12\bigl(\nabla \boldsymbol{u} +(\nabla\boldsymbol{u})^{\!\top}\bigr)\).

\paragraph{Data generation and training}
We generated a dataset of 3,134 samples by varying the time-dependent displacement boundary conditions $\bar{u}_x(t)$ and $\bar{u}_y(t)$ applied at the right edge of the domain. Each sample consists of the input displacement signals and the corresponding displacement field evolution $\mathbf{u}(\mathbf{x},t) = (u_x, u_y)^{\top}$ over the time interval, computed using the finite element method with evenly spaced time increments. The dataset was split into 2,507 training samples and 627 testing samples. The NCDE-DeepONet was trained to learn the mapping from the time-varying boundary displacements to the full spatiotemporal evolution of both displacement components.

\begin{figure}[!htb]
    \centering
    \includegraphics[width=1.0\textwidth]{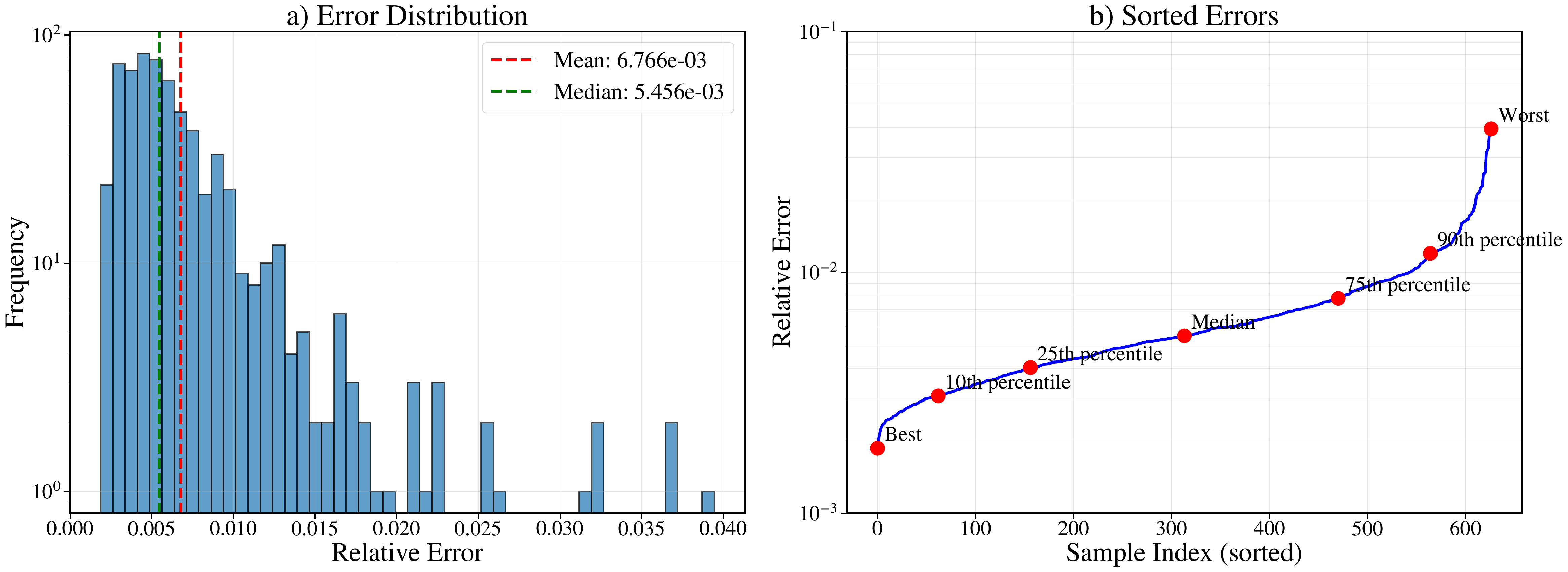}
    \caption{Error analysis of the NCDE-DeepONet model on the elastodynamics test dataset. (a) Distribution of relative errors showing a right-skewed profile with mean and median values indicated. (b) Sorted relative errors across all test samples with key percentiles marked, demonstrating consistent model performance.}
    \label{fig:elastodynamics_error}
\end{figure}

\paragraph{Error analysis}
Fig. \ref{fig:elastodynamics_error} presents a comprehensive error analysis of the trained model on the test dataset using the relative error metric defined in Eq. \eqref{eq:relative_error}. The error distribution (Fig. \ref{fig:elastodynamics_error}a) exhibits a right-skewed profile with a mean relative error of $6.77 \times 10^{-3}$ and a median of $5.46 \times 10^{-3}$. This concentration of errors at lower values indicates consistent model performance across the majority of test cases. The sorted error plot (Fig. \ref{fig:elastodynamics_error}b) reveals that the model achieves high accuracy for most samples, with relative errors below $10^{-2}$ for approximately 85\% of the test cases. The worst-case error remains bounded below $10^{-1}$, demonstrating robust generalization without significant outliers.

\begin{figure}[!htb]
    \centering
    \includegraphics[width=0.9\textwidth]{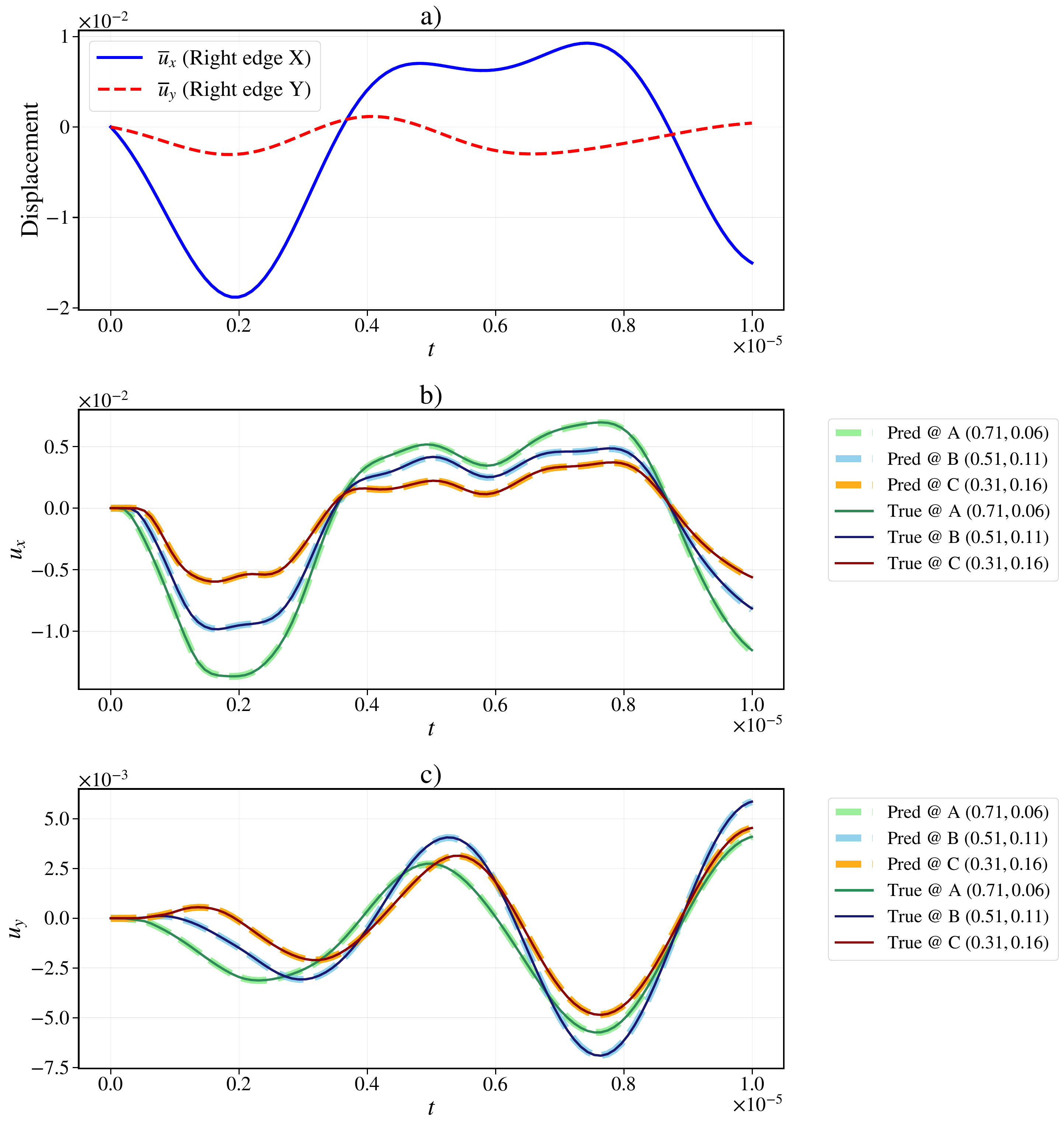}
    \caption{Time evolution of elastodynamics fields for the median error test case. a): Input displacement boundary conditions $\bar{u}_x(t)$ and $\bar{u}_y(t)$. b): True (solid lines) and predicted (dashed lines) $u_x$ at three spatial locations. c): True (solid lines) and predicted (dashed lines) $u_y$ at three spatial locations.}
    \label{fig:elastodynamics_time_evolution}
\end{figure}

\paragraph{Field predictions} To illustrate the model's predictive capabilities, we analyze the performance across different error percentiles. The corresponding input signals $\bar{u}_x(t)$ and $\bar{u}_y(t)$ for various error percentiles, showcasing diverse temporal characteristics, are provided in Fig. \ref{fig:elastodynamics_input_signals} in Appendix A.3. Fig. \ref{fig:elastodynamics_heatmap_median} shows the spatial field distribution for the median error case at the final time step. The NCDE-DeepONet accurately captures both displacement components, with the predicted fields closely matching the reference solutions. The most significant discrepancies occur near the right boundary where the time-varying displacements are imposed, consistent with the expected behavior in dynamic problems. To further demonstrate the model's temporal accuracy, Fig. \ref{fig:elastodynamics_time_evolution} presents the time evolution of the displacement fields for the median error case. The model predictions (dashed lines) closely track the reference solutions (solid lines) throughout the entire time interval at three representative spatial locations. The excellent agreement across both spatial and temporal domains confirms the model's ability to learn the complex elastodynamic response, with absolute errors for both displacement components remaining approximately two orders of magnitude smaller than their respective field values. Additional visualizations for other error percentiles, including the best, worst, and intermediate cases, are presented in Appendix A.3.

\begin{figure}[!htb]
    \centering
    \includegraphics[width=\textwidth]{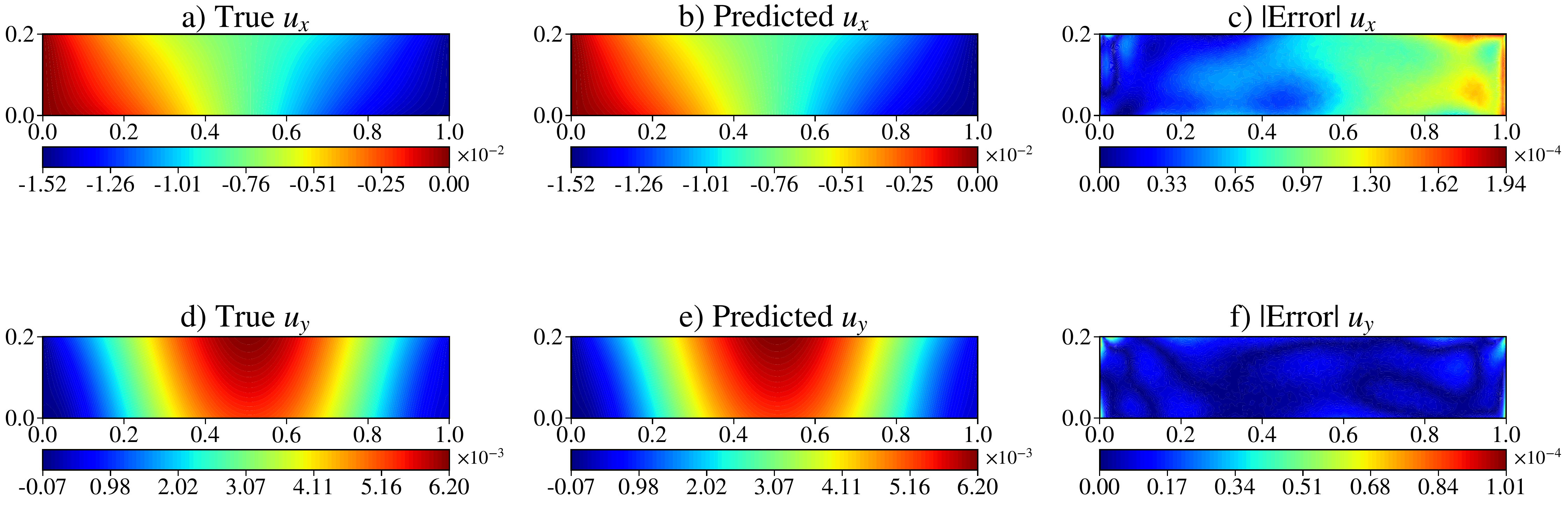}
    \caption{Elastodynamics field predictions at $t = 10^{-5}$s for the median error test case. Columns show true fields, NCDE-DeepONet predictions, and absolute errors for displacement components $u_x$ and $u_y$.}
    \label{fig:elastodynamics_heatmap_median}
\end{figure}

\begin{figure}[!htb]
    \centering
    \includegraphics[width=1.0\textwidth]{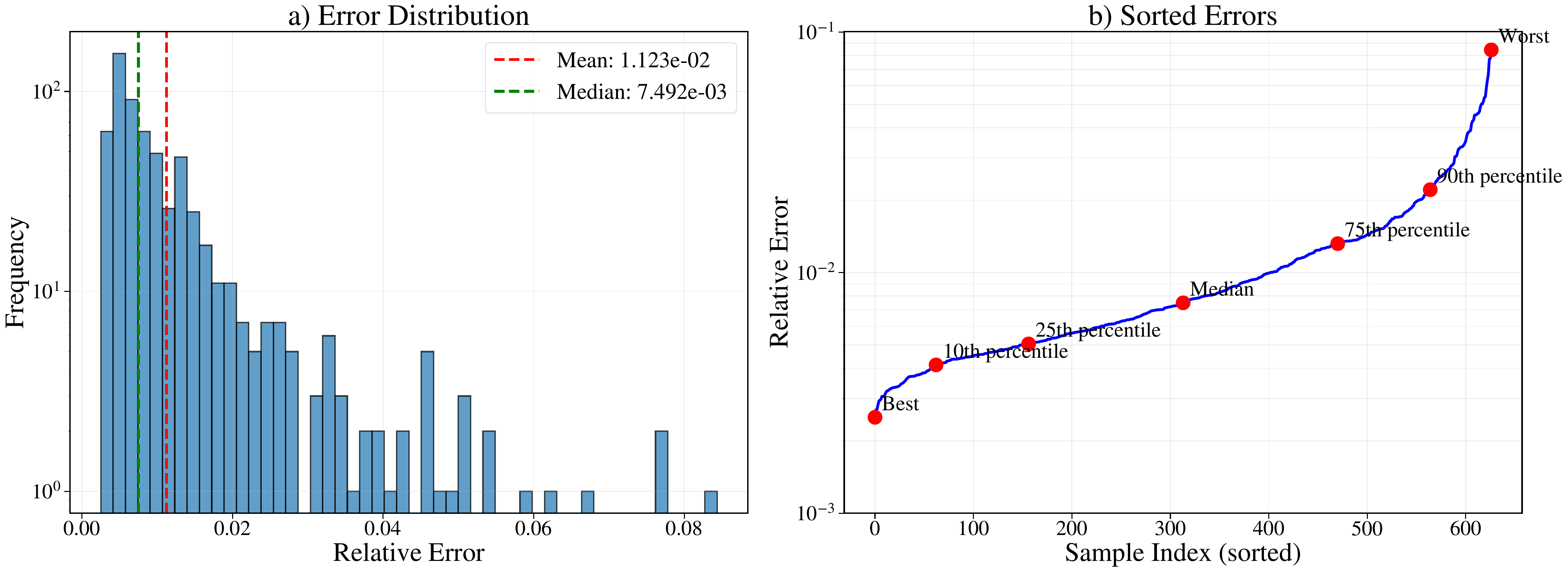}
    \caption{Error analysis of the GRU-DeepONet model on the elastodynamics test dataset. (a) Distribution of relative errors showing a right-skewed profile with higher mean and median values compared to NCDE-DeepONet. (b) Sorted relative errors across all test samples with key percentiles marked.}
    \label{fig:elastodynamics_gru_error}
\end{figure}

\paragraph{Comparison with GRU-DeepONet}
To evaluate the effectiveness of the Neural CDE architecture for encoding time-dependent boundary conditions, we compare our NCDE-DeepONet against a GRU-DeepONet baseline. The GRU variant replaces the Neural CDE branch network with a Gated Recurrent Unit (GRU) while maintaining an identical trunk network architecture. Both models were trained on the same dataset, using comparable network sizes and computational resources. For a fair comparison, both models were trained and evaluated using a fixed temporal discretization of the input signals, as the GRU-based branch network cannot handle irregularly sampled inputs—a capability unique to the NCDE-DeepONet, which will be demonstrated in the subsequent section. Fig. \ref{fig:elastodynamics_gru_error} presents the error analysis for the GRU-DeepONet model. The GRU-based architecture achieves a mean relative error of $1.12 \times 10^{-2}$ with a median of $7.49 \times 10^{-3}$, compared to the NCDE-DeepONet's superior performance of $6.77 \times 10^{-3}$ mean and $5.46 \times 10^{-3}$ median error (Fig. \ref{fig:elastodynamics_error}). This represents an approximately 40\% reduction in mean error when using the Neural CDE approach, even under conditions where both architectures receive regularly sampled inputs. Furthermore, the error distribution of the NCDE-DeepONet is more concentrated around lower values, with approximately 85\% of test cases achieving errors below $10^{-2}$, compared to only 65\% for the GRU variant. These results demonstrate that the continuous-time formulation of Neural CDEs provides a more effective representation of the temporal dynamics in elastodynamic problems, particularly for capturing the complex wave propagation patterns induced by time-varying boundary displacements.

\subsection{Thermoelasticity}\label{thermoelasticity}

In the small-strain, linear thermoelastic regime—where inertial effects are neglected and all body forces and internal heat sources vanish—the coupled fields \(\boldsymbol{u}(\boldsymbol{x},t)\) and \(T(\boldsymbol{x},t)\) satisfy the quasi-static system  
\begin{subequations}\label{eq:thermo_pdes}
\begin{align}
\boldsymbol{\nabla}\!\cdot\!\boldsymbol{\sigma} &= \mathbf 0,
      &&\boldsymbol{x}\in\Omega,\; t\in(0,T], \label{eq:blm}\\
\rho T_{0}\,\dot{\eta}\;+\;\boldsymbol{\nabla}\!\cdot\!\boldsymbol{q} &= 0,
      &&\boldsymbol{x}\in\Omega,\; t\in(0,T], \label{eq:energy}
\end{align}
\end{subequations}
where \(\boldsymbol{\sigma}\) is the Cauchy stress, \(\boldsymbol{q}\) the heat flux, \(\eta\) the entropy per unit mass, and \(\rho\) the mass density.  Fourier’s law provides the constitutive relation for heat conduction,  
\begin{equation}
\begin{aligned}
    \boldsymbol{q} = -k \boldsymbol{\nabla} T,
    \label{FL}
\end{aligned}
\end{equation}
with thermal conductivity \(k\).  Under the infinitesimal-strain assumption, the kinematics reduce to $\boldsymbol{\varepsilon}=\tfrac12\bigl(\boldsymbol{\nabla}\boldsymbol{u}+\boldsymbol{\nabla}\boldsymbol{u}^{\!\top}\bigr)$, and the isotropic plane-stress thermoelastic law reads  
\begin{subequations}\label{eq:thermo_constitutive}
\begin{align}
\boldsymbol{\sigma} &=
      2\mu\,\boldsymbol{\varepsilon}
      +\bigl[\lambda\,\operatorname{tr}(\boldsymbol{\varepsilon})
            \;-\;\alpha\,(3\lambda+2\mu)\,(T-T_{0})\bigr]\mathbf I,
      \label{eq:sigma}\\
\rho T_{0}\,\dot{\eta} &=
      \rho C_{\varepsilon}\,\dot{T}
      +\alpha\,(3\lambda+2\mu)\,T_{0}\,\operatorname{tr}(\dot{\boldsymbol{\varepsilon}}),
      \label{eq:entropy}
\end{align}
\end{subequations}
where \(C_{\varepsilon}\) is the specific heat at constant strain,
\(\alpha\) the coefficient of thermal expansion, and \(\lambda,\mu\) are the Lamé parameters.  Eliminating \(\dot{\eta}\) in
\eqref{eq:energy} by means of \eqref{eq:entropy} yields the compact energy balance
\begin{equation}
\begin{aligned}
\rho C_{\varepsilon}\,\dot{T}
+\alpha\,(3\lambda+2\mu)\,T_{0}\,\operatorname{tr}(\dot{\boldsymbol{\varepsilon}})
+\boldsymbol{\nabla}\!\cdot\!\boldsymbol{q}=0.
\end{aligned}
\end{equation}

\begin{figure}[!htb]
    \centering
    \includegraphics[width=0.6\textwidth]{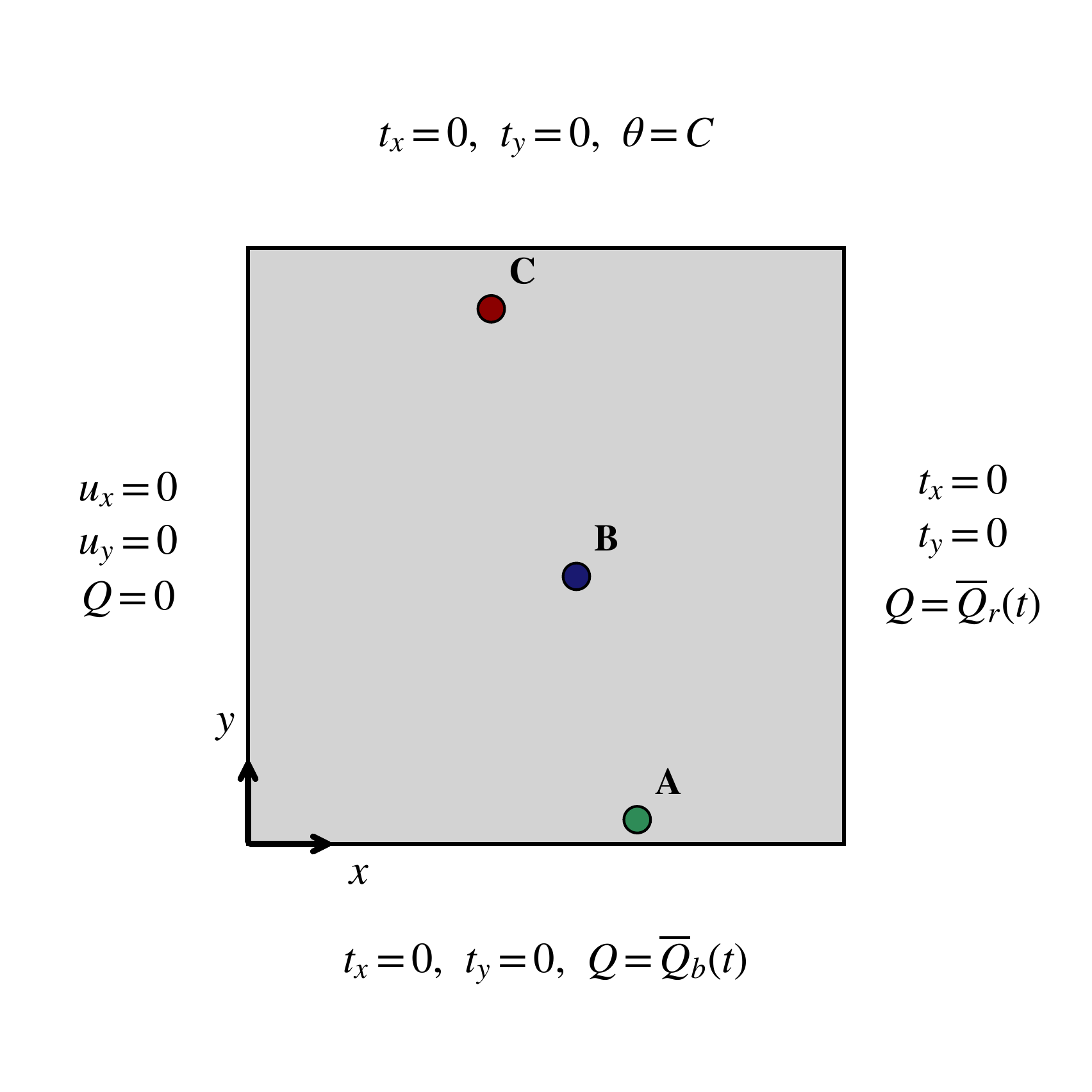}
    \caption{Schematic of the geometry and boundary conditions of the thermoelasticity example.}
    \label{Figure_Schematic_thermo}
\end{figure}

For this problem, we consider a rectangular domain with mixed thermal and mechanical boundary conditions. The left edge is mechanically fixed ($u_x = u_y = 0$) and thermally insulated ($Q = 0$). The top edge is stress-free ($t_x = t_y = 0$) with a prescribed constant temperature ($\theta = 10$). The right and bottom edges are stress-free and subjected to time-varying heat fluxes $\bar{Q}_r(t)$ and $\bar{Q}_b(t)$, respectively. This configuration creates a complex thermoelastic response where thermal loading induces mechanical deformations through the coupling terms in the constitutive equations.

\paragraph{Data generation and training}
We generated a dataset of 5,000 samples by varying the time-dependent heat flux profiles $\bar{Q}_r(t)$ and $\bar{Q}_b(t)$ applied at the right and bottom boundaries, respectively. Each sample consists of the input heat flux signals and the corresponding coupled displacement-temperature field evolution over the time interval, computed using the finite element method with evenly spaced time increments. The dataset was split into 4,000 samples for training and 1,000 samples for testing. The NCDE-DeepONet was trained to learn the mapping from the time-varying boundary conditions to the full spatiotemporal evolution of the displacement components $(u_x, u_y)$ and temperature field $\theta$.

\begin{figure}[!htb]
    \centering
    \includegraphics[width=1.0\textwidth]{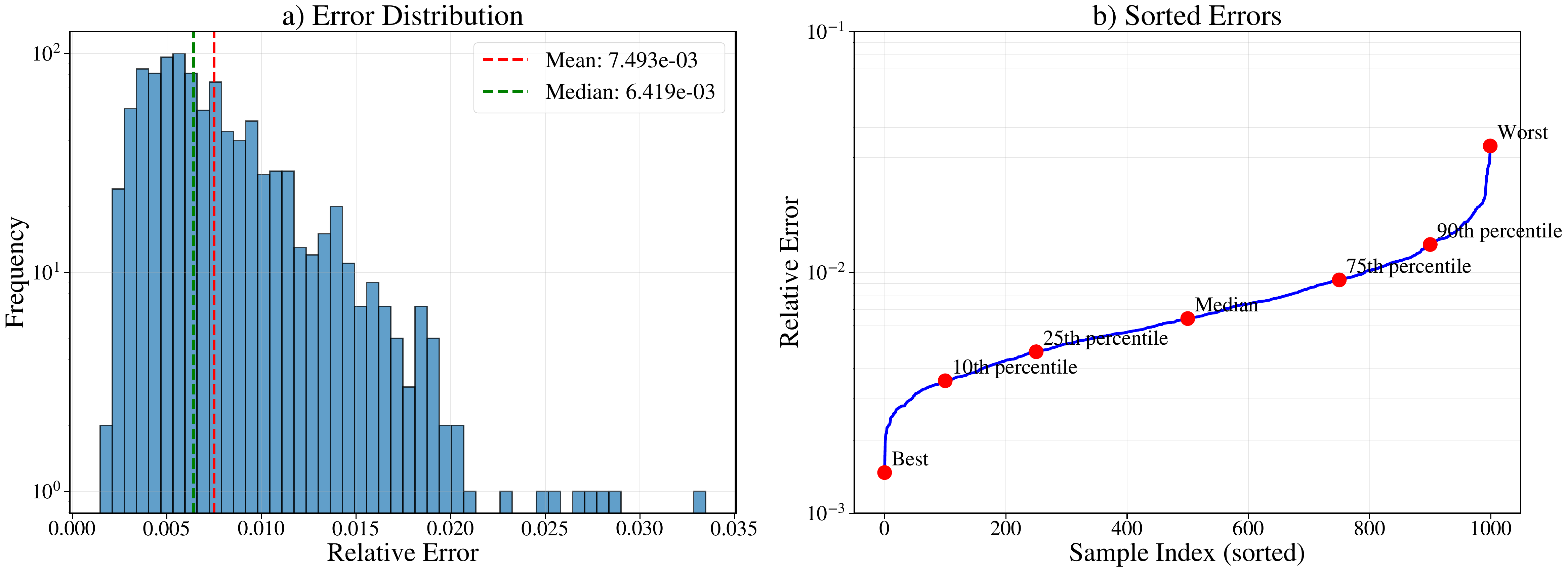}
    \caption{Error analysis of the NCDE-DeepONet model on the thermoelasticity test dataset. (a) Distribution of relative errors showing a right-skewed profile with mean and median values indicated. (b) Sorted relative errors across all test samples with key percentiles marked, demonstrating consistent model performance.}
    \label{fig:thermoelasticity_error}
\end{figure}

\paragraph{Error analysis}
Fig. \ref{fig:thermoelasticity_error} presents a comprehensive error analysis of the trained model on the test dataset using the relative error metric defined in Eq. \eqref{eq:relative_error}. The error distribution (Fig. \ref{fig:thermoelasticity_error}a) exhibits a right-skewed profile with a mean relative error of $7.49 \times 10^{-3}$ and a median of $6.42 \times 10^{-3}$. The sorted error plot (Fig. \ref{fig:thermoelasticity_error}b) reveals that the model achieves consistent accuracy across the majority of test cases, with relative errors below $10^{-2}$ for approximately 90\% of the samples. The worst-case error remains bounded below $10^{-1}$, demonstrating robust generalization performance.

\begin{figure}[!htb]
    \centering
    \includegraphics[width=\textwidth]{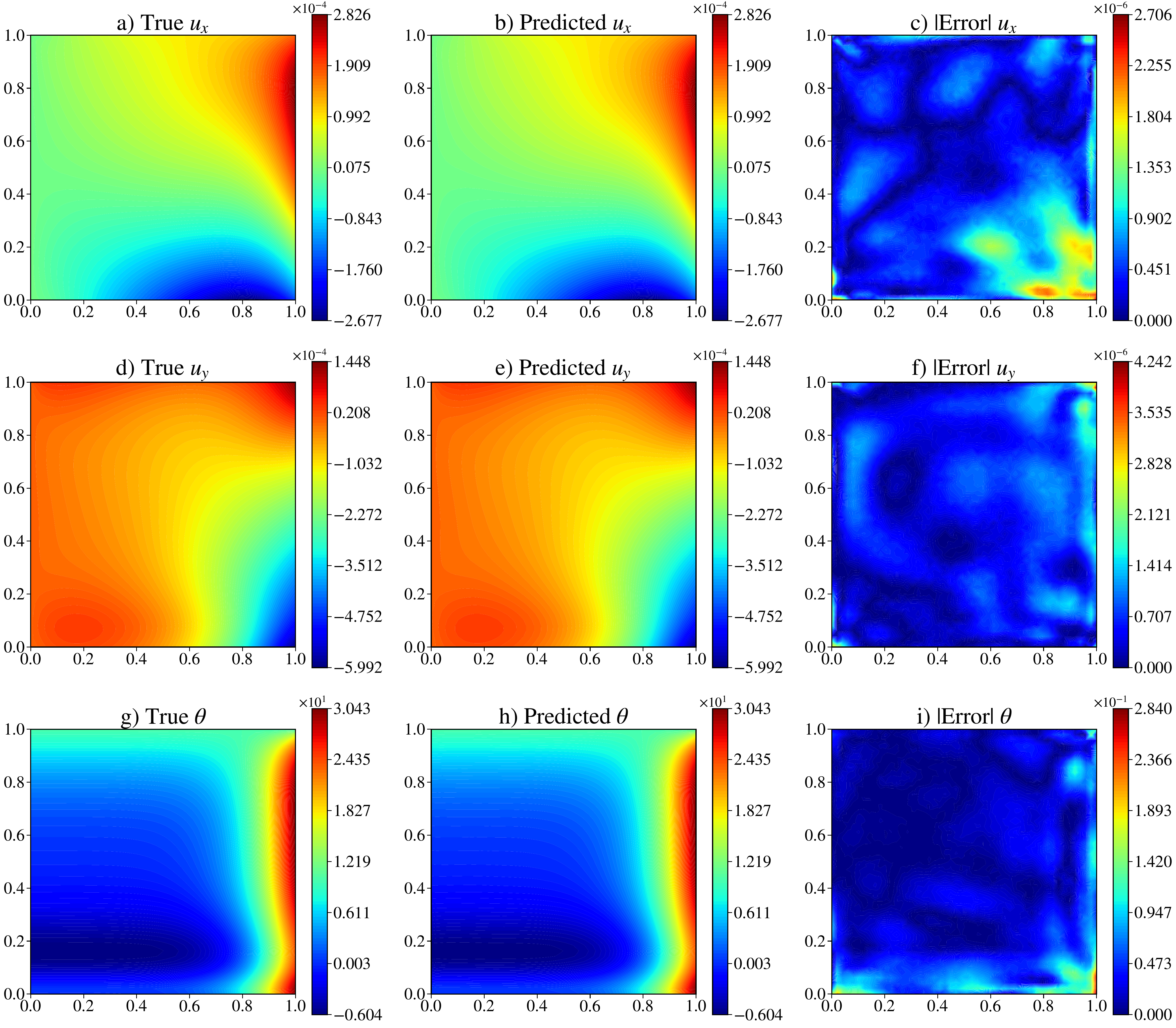}
    \caption{Thermoelasticity field predictions at $t = 250$s for the median error test case. Columns show true fields, NCDE-DeepONet predictions, and absolute errors for displacement components $u_x$, $u_y$, and temperature $\theta$.}
    \label{fig:thermoelasticity_heatmap_t49}
\end{figure}

\begin{figure}[!htb]
    \centering
    \includegraphics[width=\textwidth]{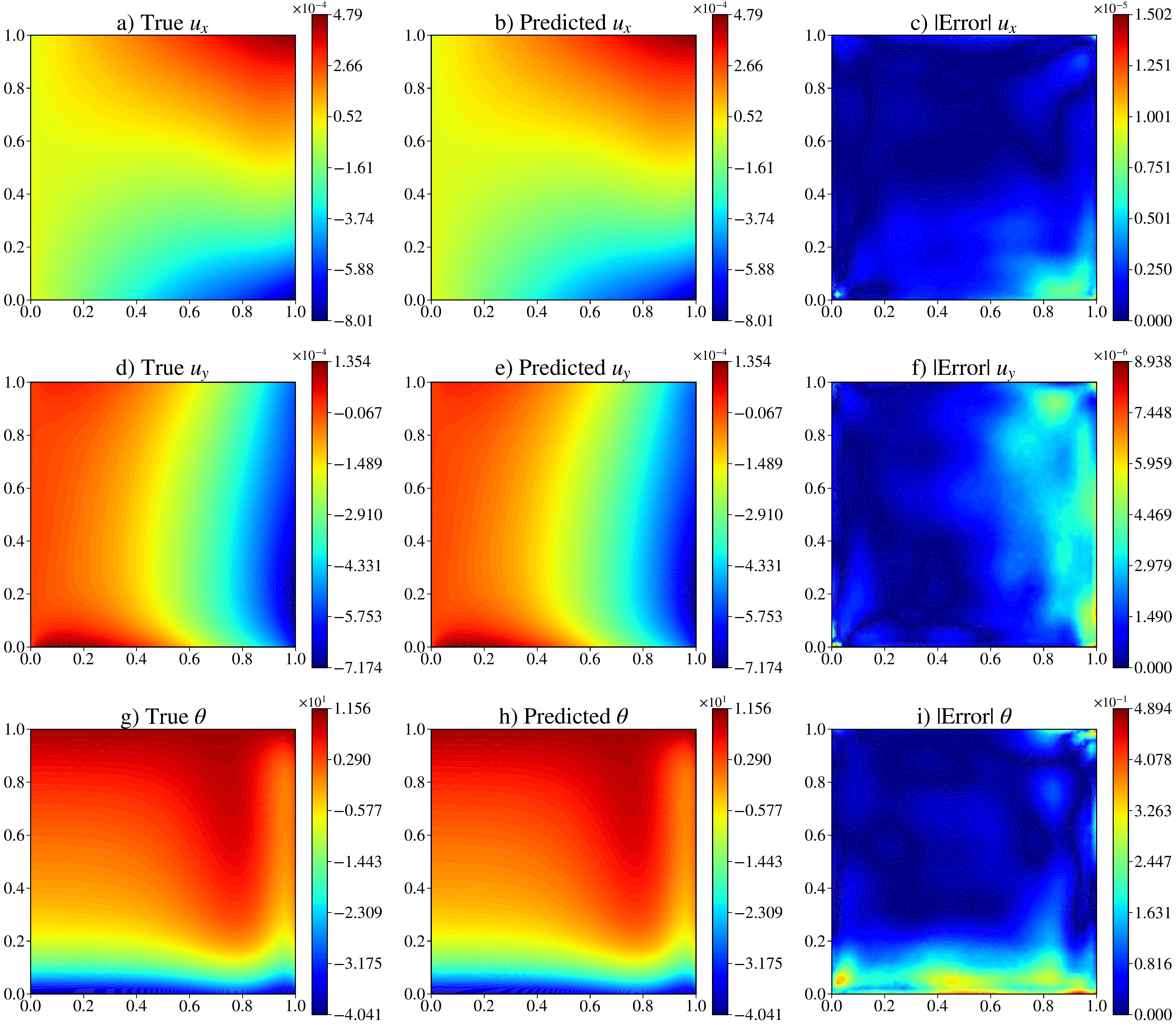}
    \caption{Thermoelasticity field predictions at $t = 500$s for the median error test case. The model maintains accuracy throughout the time evolution, with errors concentrated near the flux boundaries.}
    \label{fig:thermoelasticity_heatmap_t98}
\end{figure}

\begin{figure}[!htb]
    \centering
    \includegraphics[width=0.9\textwidth]{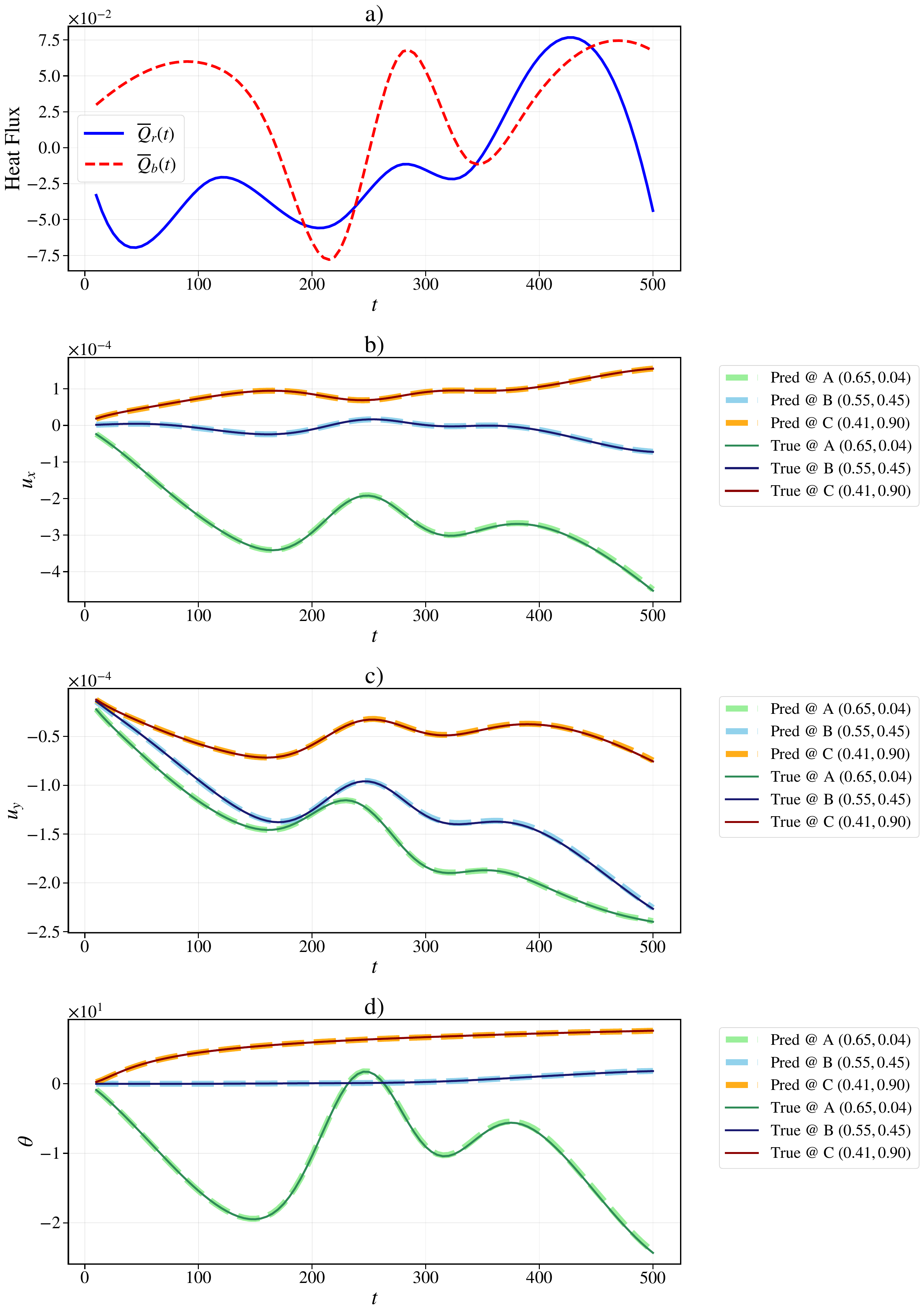}
    \caption{Time evolution of thermoelasticity fields for the median error test case. Top: Input heat flux boundary conditions $\bar{Q}_r(t)$ and $\bar{Q}_b(t)$. Bottom panels: True (solid lines) and predicted (dashed lines) values of displacement components $u_x$, $u_y$, and temperature $\theta$ at three spatial locations.}
    \label{fig:thermoelasticity_time_evolution}
\end{figure}

\paragraph{Field predictions}
To illustrate the model's predictive capabilities, Figs. \ref{fig:thermoelasticity_heatmap_t49} and \ref{fig:thermoelasticity_heatmap_t98} show the true fields, predictions, and absolute errors for all three components $(u_x, u_y, \theta)$ at two time instances for the median error case, while Fig. \ref{fig:thermoelasticity_time_evolution} presents the temporal evolution at three representative spatial locations. The corresponding input heat flux signals $\bar{Q}_r(t)$ and $\bar{Q}_b(t)$ for this case, along with signals for other error percentiles, are shown in Fig. \ref{fig:thermoelasticity_input_signals} in Appendix A.4. The results demonstrate that the NCDE-DeepONet accurately captures both the spatiotemporal evolution and the coupled nature of the thermoelastic response. In the temporal domain (Fig. \ref{fig:thermoelasticity_time_evolution}), the predicted trajectories closely track the reference solutions across all field components and monitored locations, indicating the model's ability to learn the complex dynamics induced by the time-varying thermal loads. The spatial field predictions (Figs. \ref{fig:thermoelasticity_heatmap_t49} and \ref{fig:thermoelasticity_heatmap_t98}) reveal that the model correctly reproduces the expected physical behavior: mechanical deformations arising from thermal expansion and heat diffusion from the flux boundaries. While the absolute errors are largest near the heat flux boundaries—a challenging region due to the imposed boundary conditions—these discrepancies remain within acceptable bounds across the domain. This consistent performance in both space and time validates the NCDE-DeepONet's effectiveness in learning the solution operator for coupled multiphysics problems. Additional visualizations, including field predictions at different error percentiles and complete time evolution plots, are provided in Appendix A.4.

\paragraph{Interpolation capability of NCDE-DeepONet}
A key advantage of employing Neural CDEs in the branch network is their inherent ability to handle variable input signal sampling rates. Although the model was trained exclusively with fixed time increments ($\sim$99 evenly spaced points), the continuous-time formulation enables it to process boundary conditions with arbitrary temporal resolution. To demonstrate this capability, we evaluated the trained model with three different input scenarios: the original sampling (100\%), a sparse sampling with only half the data points (50\%), and an upsampled version with double the temporal resolution (200\%).

\begin{figure}[!htb]
    \centering
    \includegraphics[width=\textwidth]{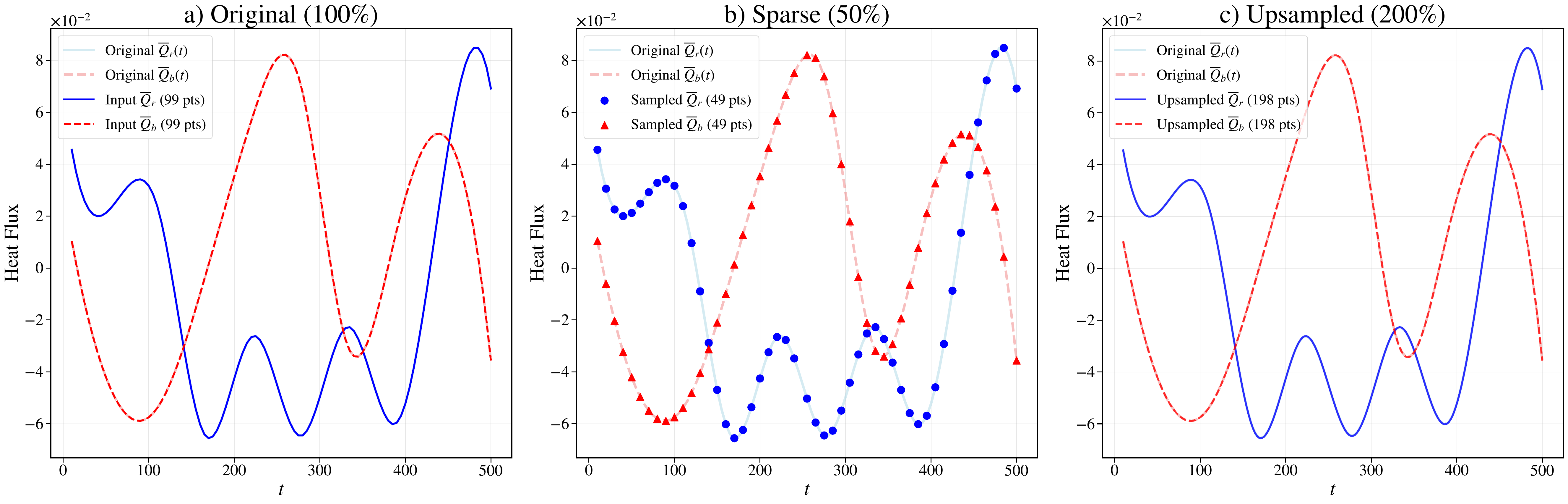}
    \caption{Input heat flux boundary conditions under different sampling scenarios. Left: Original sampling with 99 points as used during training. Center: Sparse sampling with only 49 points (50\%), requiring the NCDE to interpolate between observations. Right: Upsampled to 198 points (200\%), demonstrating the model's ability to process denser temporal information.}
    \label{fig:ncde_input_sampling}
\end{figure}

Fig. \ref{fig:ncde_input_sampling} illustrates these three input scenarios for the heat flux boundary conditions $\bar{Q}_r(t)$ and $\bar{Q}_b(t)$. In the sparse sampling case (center panel), only 49 discrete observations are provided, yet the NCDE internally interpolates between these points through its learned dynamics. Conversely, the upsampled case (right panel) provides 198 data points, offering a denser representation of the same underlying signals. Despite these significant variations in input sampling, ranging from sparse discrete points to densely sampled curves, the model can predict the solution fields at any spatiotemporal points as dictated by the trunk.

\begin{figure}[!htb]
    \centering
    \includegraphics[width=\textwidth]{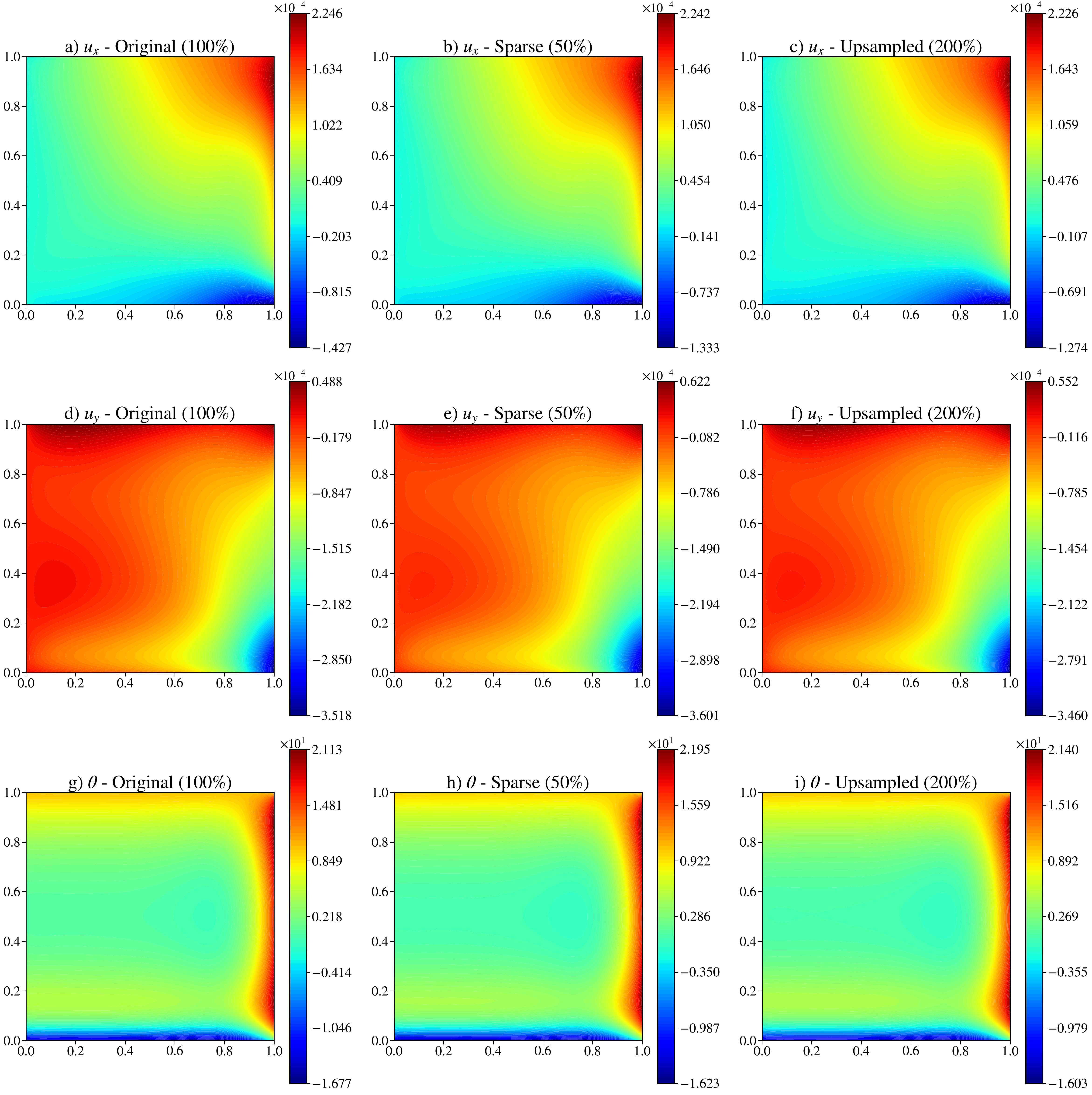}
    \caption{Spatial field predictions at $t = 250$s for different input sampling scenarios. Despite receiving boundary conditions with 50\%, 100\%, or 200\% of the original temporal resolution, the NCDE-DeepONet produces consistent spatial fields for displacement components $(u_x, u_y)$ and temperature $\theta$.}
    \label{fig:ncde_spatial_fields_comparison}
\end{figure}

Fig. \ref{fig:ncde_spatial_fields_comparison} presents the predicted spatial fields for all three components $(u_x, u_y, \theta)$ at $t = 250$s under these different input sampling scenarios. Remarkably, the spatial field predictions remain in agreement across all sampling rates, with the temperature and displacement patterns showing consistent magnitudes and distributions. This demonstrates that the NCDE effectively learns a continuous representation of the input signals, enabling it to maintain prediction accuracy regardless of the temporal discretization of the boundary conditions (i.e., the input signals).

This interpolation capability has significant practical implications. In real-world applications, boundary condition measurements may come from sensors with varying sampling rates, experience data dropouts, or require upsampling for analysis. The NCDE-DeepONet's ability to seamlessly handle such variations without retraining makes it particularly suitable for deployment in scenarios where data quality and sampling rates cannot be guaranteed to match the training conditions. This flexibility, combined with the model's demonstrated accuracy in learning complex thermoelastic coupling, validates the choice of NCDEs for encoding time-dependent boundary conditions in multiphysics problems.
\section{Conclusions, limitations, and future work}\label{conclu}

This work advances data-driven modeling of transient mechanics by introducing NCDE-DeepONet, which fuses a Neural Controlled Differential Equation branch with a spatio-temporal trunk to learn solution operators directly from irregular load histories. By embedding continuous-time dynamics within the operator framework, the proposed architecture addresses the limitations of Neural ODEs in incorporating new time-varying inputs and the step-size sensitivity of recurrent DeepONets, thereby providing a principled, mesh-independent surrogate for complex PDE systems. 

Our numerical experiments across three benchmark problems—transient Poisson, elastodynamics, and thermoelasticity—demonstrate the framework's accuracy and versatility. The model consistently achieves relative errors below $10^{-2}$ for the majority of test cases. Notably, the NCDE-DeepONet outperforms the GRU-DeepONet baseline by approximately $40\%$ in mean error, even under regular sampling conditions where both architectures should perform comparably. More significantly, the continuous-time formulation enables seamless handling of variable input sampling rates, as evidenced by consistent predictions when tested with $50\%$ sparse and $200\%$ upsampled input signals—a capability unattainable with discrete-time approaches.

The practical implications extend beyond numerical accuracy. In real-world applications where sensor data arrives irregularly, experiences dropouts, or requires temporal upsampling, the NCDE-DeepONet can adapt without retraining. The spatiotemporal trunk further enables continuous queries across space and time, eliminating the need for post-processing interpolation and providing true resolution independence for downstream analysis. Future work could explore extensions to three-dimensional domains, investigate the framework's performance under extreme extrapolation scenarios, and develop theoretical guarantees for approximation accuracy. As computational mechanics increasingly embraces data-driven methods, the NCDE-DeepONet represents a step toward truly flexible, continuous-time operator learning.

\section*{Acknowledgment}
This work was partially supported by the Sand Hazards and Opportunities for Resilience, Energy, and Sustainability (SHORES) Center, funded by Tamkeen under the NYUAD Research Institute Award CG013. The authors wish to thank the NYUAD Center for Research Computing for providing resources, services, and skilled personnel. This work was partially supported by Sandooq Al Watan Applied Research and Development (SWARD), funded by Grant No.: SWARD-F22-018.
\section*{Data availability}
The data supporting the study's findings will be available upon paper acceptance.

\bibliography{mybibfile}
\appendix
\setcounter{table}{0}
\renewcommand{\thetable}{A\arabic{table}}
\setcounter{figure}{0}
\renewcommand{\thefigure}{A\arabic{figure}}

\section*{Appendix: Supporting Information}

\subsection*{A.1. Hyperparameters and network sizes}

The NCDE-DeepONet architecture employs consistent hyperparameters across all three benchmark problems (transient Poisson, elastodynamics, and thermoelasticity), demonstrating the framework's robustness without requiring problem-specific tuning. The network architecture consists of a Neural CDE branch and an MLP trunk, both configured with identical depth and width for balanced representation capacity. The branch network processes time-dependent boundary conditions through a controlled differential equation, while the trunk network encodes spatiotemporal coordinates.

Training follows a warmup cosine decay schedule for the learning rate, enabling stable convergence from an initial exploration phase to final refinement. The ODE solver configuration strikes a balance between accuracy and computational efficiency through adaptive step-size control with specified error tolerances. Cubic interpolation constructs continuous control paths from discrete input observations, providing smooth derivatives for the Neural CDE dynamics. All experiments use the same random seed to ensure reproducibility. Table \ref{tab:hyperparameters} summarizes the complete hyperparameter configuration.

\begin{table}[htbp]
\centering
\caption{Hyperparameter configuration used for all numerical experiments}
\label{tab:hyperparameters}
\begin{tabular}{lll}
\toprule
\textbf{Category} & \textbf{Parameter} & \textbf{Value} \\
\midrule
\multirow{4}{*}{Network Architecture} 
    & Branch hidden size & 200 \\
    & Branch layers & 6 \\
    & Trunk hidden size & 200 \\
    & Trunk layers & 6 \\
\midrule
\multirow{5}{*}{Training} 
    & Batch size & 128 \\
    & Epochs & 4000 \\
    & Initial learning rate & $10^{-3}$ \\
    & Final learning rate & $10^{-5}$ \\
    & LR scheduler & Warmup cosine decay \\
\midrule
\multirow{5}{*}{ODE Solver} 
    & Solver type & Tsit5 \\
    & Relative tolerance & $10^{-4}$ \\
    & Absolute tolerance & $10^{-7}$ \\
    & Maximum steps & 50,000 \\
    & Interpolation & Cubic Hermite \\
\midrule
\multirow{2}{*}{Other} 
    & Random seed & 2024 \\
    & Normalization & maximum absolute scaling \\
\bottomrule
\end{tabular}
\end{table}

While the network architecture hyperparameters remain fixed across all problems, the total number of trainable parameters varies due to differences in input and output dimensionality. The branch network's input dimension depends on the number of time-dependent boundary conditions (one for Poisson, two for elastodynamics, and two for thermoelasticity), while the output dimension is determined by the number of field components (one scalar field for Poisson, two displacement components for elastodynamics, and three fields for thermoelasticity including temperature). These dimensional differences affect the first layer of the branch network and the final projection layer that maps branch outputs to field components, resulting in total parameter counts of approximately 609K for Poisson, 690K for elastodynamics, and 730K for thermoelasticity. Despite this variation, the core network capacity—determined by the hidden layer dimensions and depth—remains consistent.

\subsection*{A.2. Additional results: Poisson}

\begin{figure}[!htb]
    \centering
    \includegraphics[width=1.0\textwidth]{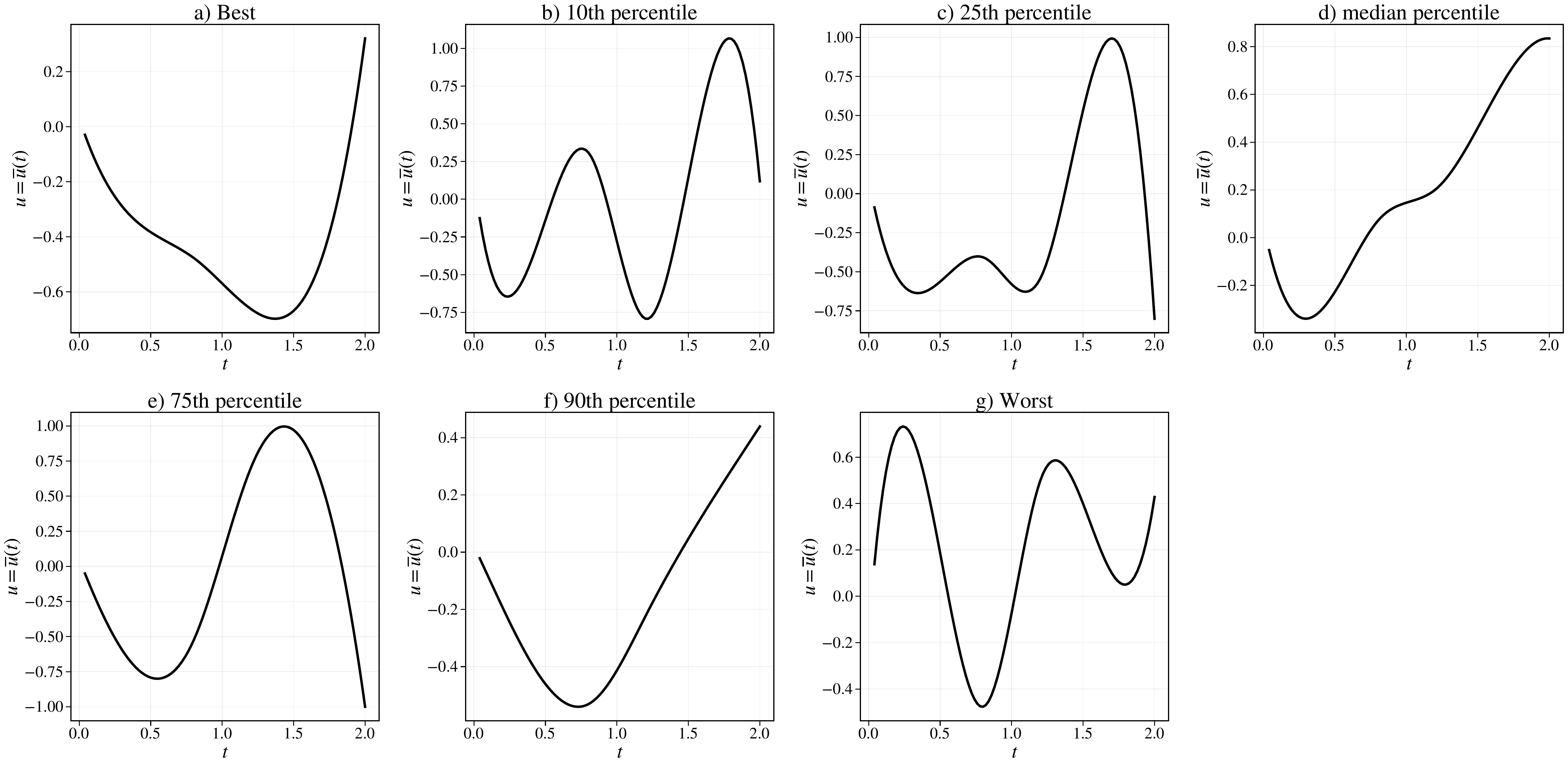}
    \caption{Input boundary conditions $\overline{u}(t)$ for the Poisson test cases at different error percentiles. The signals demonstrate the diversity of temporal profiles in the test dataset.}
    \label{fig:poisson_input_signals}
\end{figure}

This section provides supplementary visualizations for the Poisson problem discussed in Section \ref{poisson}. Figure \ref{fig:poisson_input_signals} displays the input Dirichlet boundary conditions $\overline{u}(t)$ for test cases spanning the entire error spectrum, from best to worst performance. These signals exhibit diverse temporal characteristics, including smooth variations, oscillatory patterns with multiple extrema, and signals with varying amplitudes and rates of change, demonstrating the range of boundary condition profiles included in the test dataset.

\begin{figure}[!htb]
    \centering
    \includegraphics[width=1.0\textwidth]{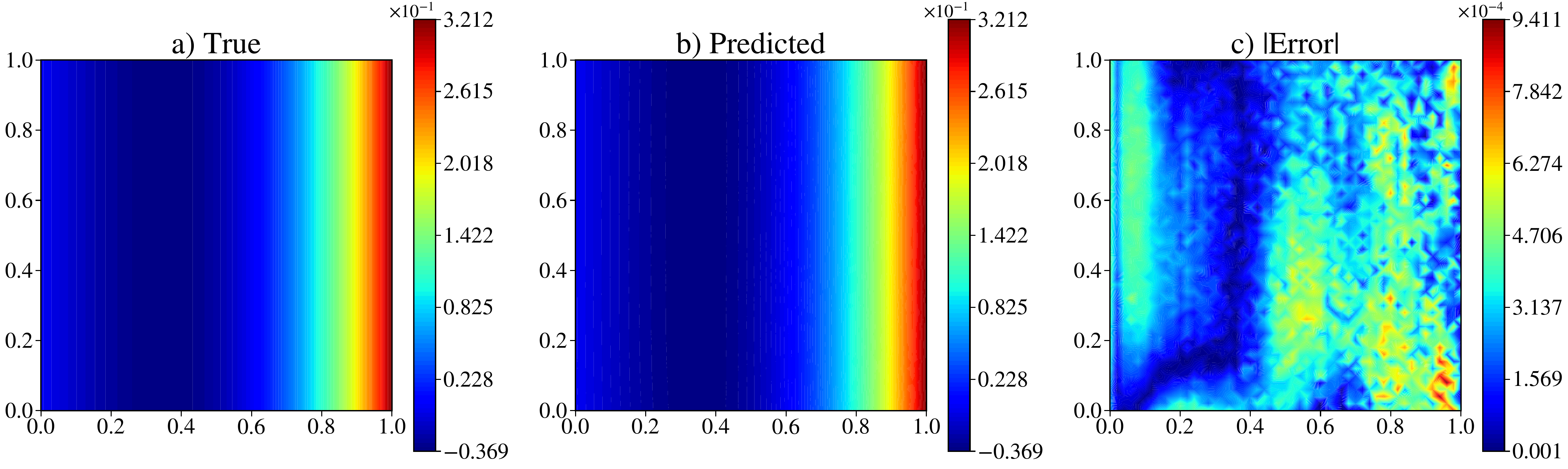}
    \caption{Poisson field predictions at $t = 2$s for the best error test case. Columns show true field, NCDE-DeepONet prediction, and absolute error for the scalar field $u(\mathbf{x},t)$.}
    \label{fig:poisson_heatmap_best}
\end{figure}

\begin{figure}[!htb]
    \centering
    \includegraphics[width=1.0\textwidth]{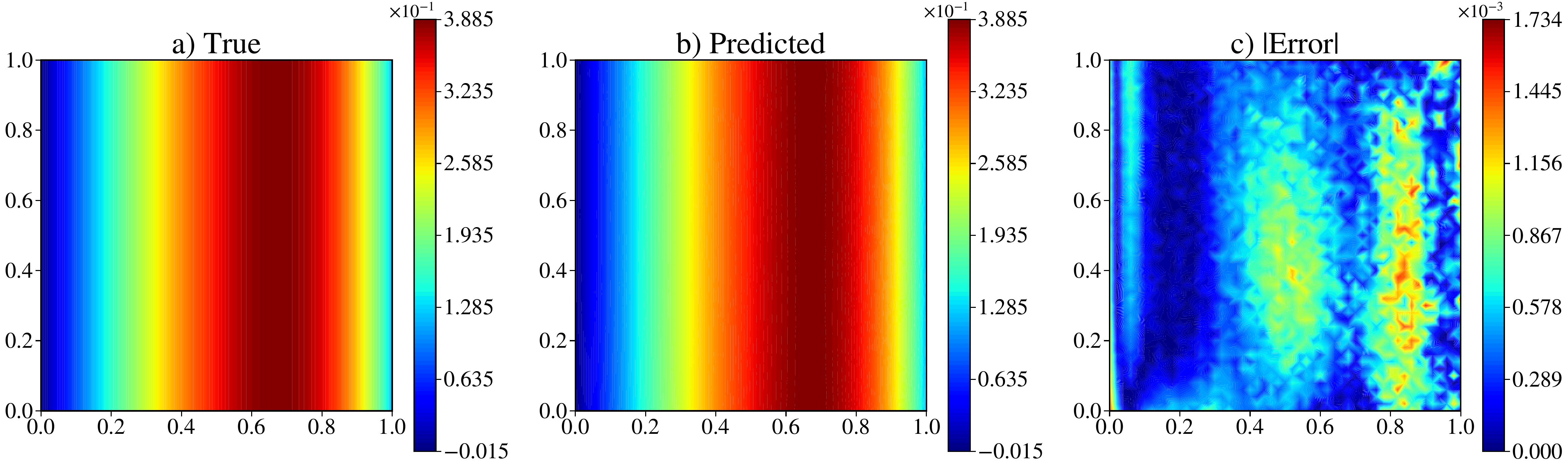}
    \caption{Poisson field predictions at $t = 2$s for the 10th percentile error test case. Columns show true field, NCDE-DeepONet prediction, and absolute error for the scalar field $u(\mathbf{x},t)$.}
    \label{fig:poisson_heatmap_10th}
\end{figure}

\begin{figure}[!htb]
    \centering
    \includegraphics[width=1.0\textwidth]{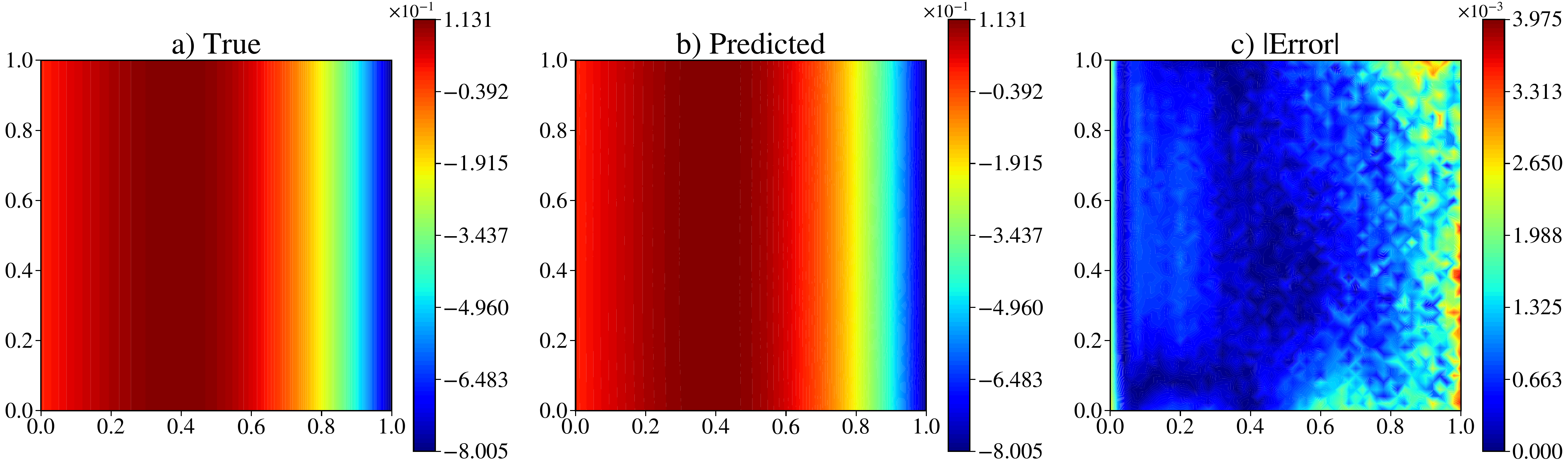}
    \caption{Poisson field predictions at $t = 2$s for the 25th percentile error test case. Columns show true field, NCDE-DeepONet prediction, and absolute error for the scalar field $u(\mathbf{x},t)$.}
    \label{fig:poisson_heatmap_25th}
\end{figure}

\begin{figure}[!htb]
    \centering
    \includegraphics[width=1.0\textwidth]{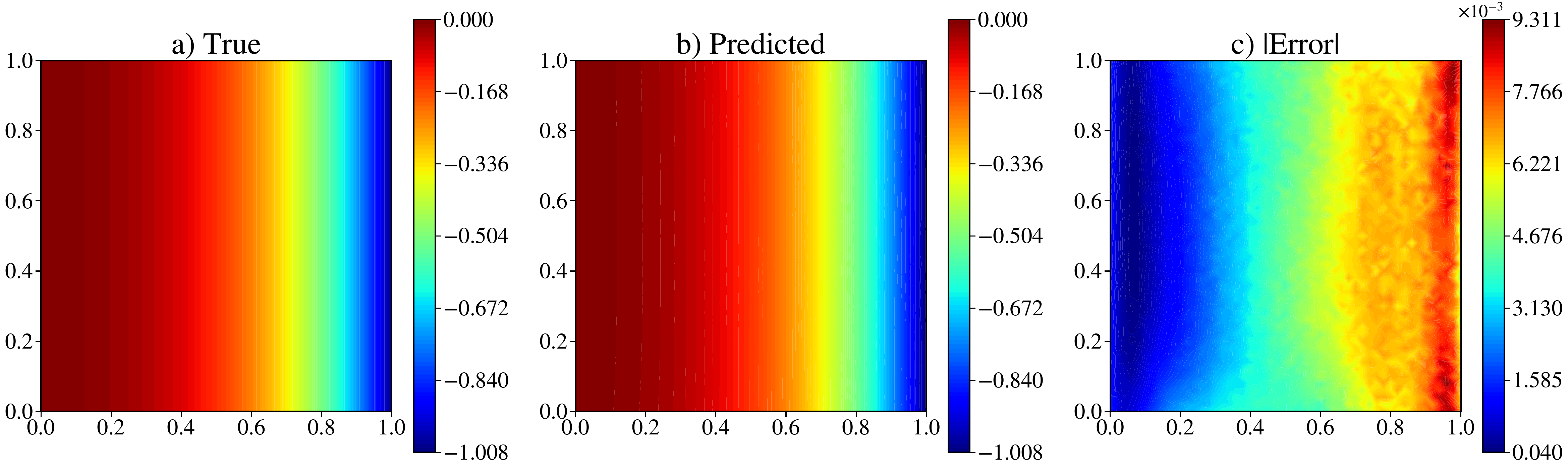}
    \caption{Poisson field predictions at $t = 2$s for the 75th percentile error test case. Columns show true field, NCDE-DeepONet prediction, and absolute error for the scalar field $u(\mathbf{x},t)$.}
    \label{fig:poisson_heatmap_75th}
\end{figure}

\begin{figure}[!htb]
    \centering
    \includegraphics[width=1.0\textwidth]{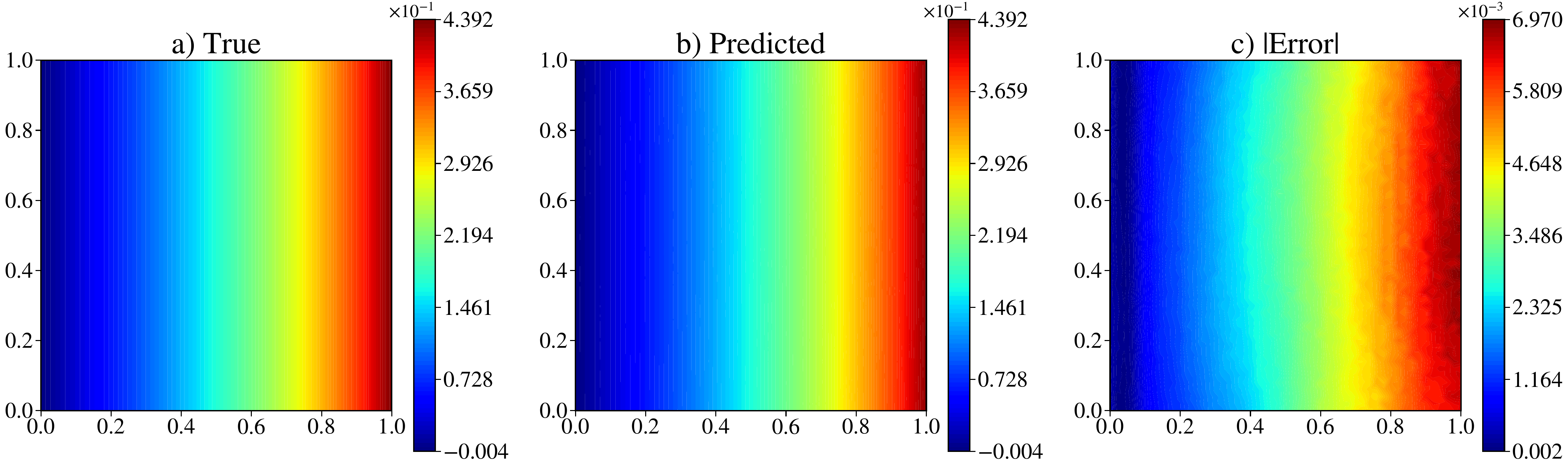}
    \caption{Poisson field predictions at $t = 2$s for the 90th percentile error test case. Columns show true field, NCDE-DeepONet prediction, and absolute error for the scalar field $u(\mathbf{x},t)$.}
    \label{fig:poisson_heatmap_90th}
\end{figure}

\begin{figure}[!htb]
    \centering
    \includegraphics[width=1.0\textwidth]{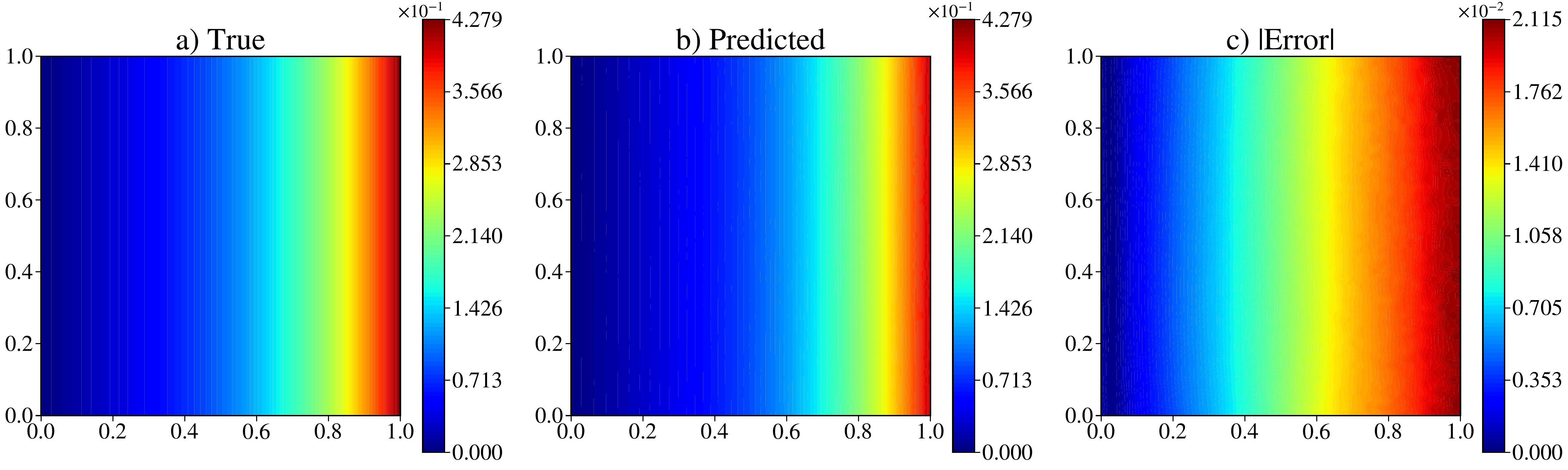}
    \caption{Poisson field predictions at $t = 2$s for the worst error test case. Columns show true field, NCDE-DeepONet prediction, and absolute error for the scalar field $u(\mathbf{x},t)$.}
    \label{fig:poisson_heatmap_worst}
\end{figure}

Figures \ref{fig:poisson_heatmap_best}--\ref{fig:poisson_heatmap_worst} present the spatial field distributions at the final time $t = 2$s for test cases at different error percentiles. These visualizations reveal several important observations. First, the model consistently captures the diffusive behavior governed by the transient Poisson equation across all error levels, maintaining the expected smooth transition from the fixed left boundary to the time-varying right boundary. Second, the error patterns show a consistent structure across percentiles, with larger discrepancies concentrated near the right boundary where the dynamic Dirichlet condition is imposed. Finally, comparing the best and worst cases demonstrates that prediction accuracy is primarily influenced by the temporal complexity of the input signal $\overline{u}(t)$, with the model successfully reproducing the fundamental physics even in challenging cases with rapid temporal variations or multiple oscillations.

\subsection*{A.3. Additional results: Elastodynamics}

\begin{figure}[!htb]
    \centering
    \includegraphics[width=\textwidth]{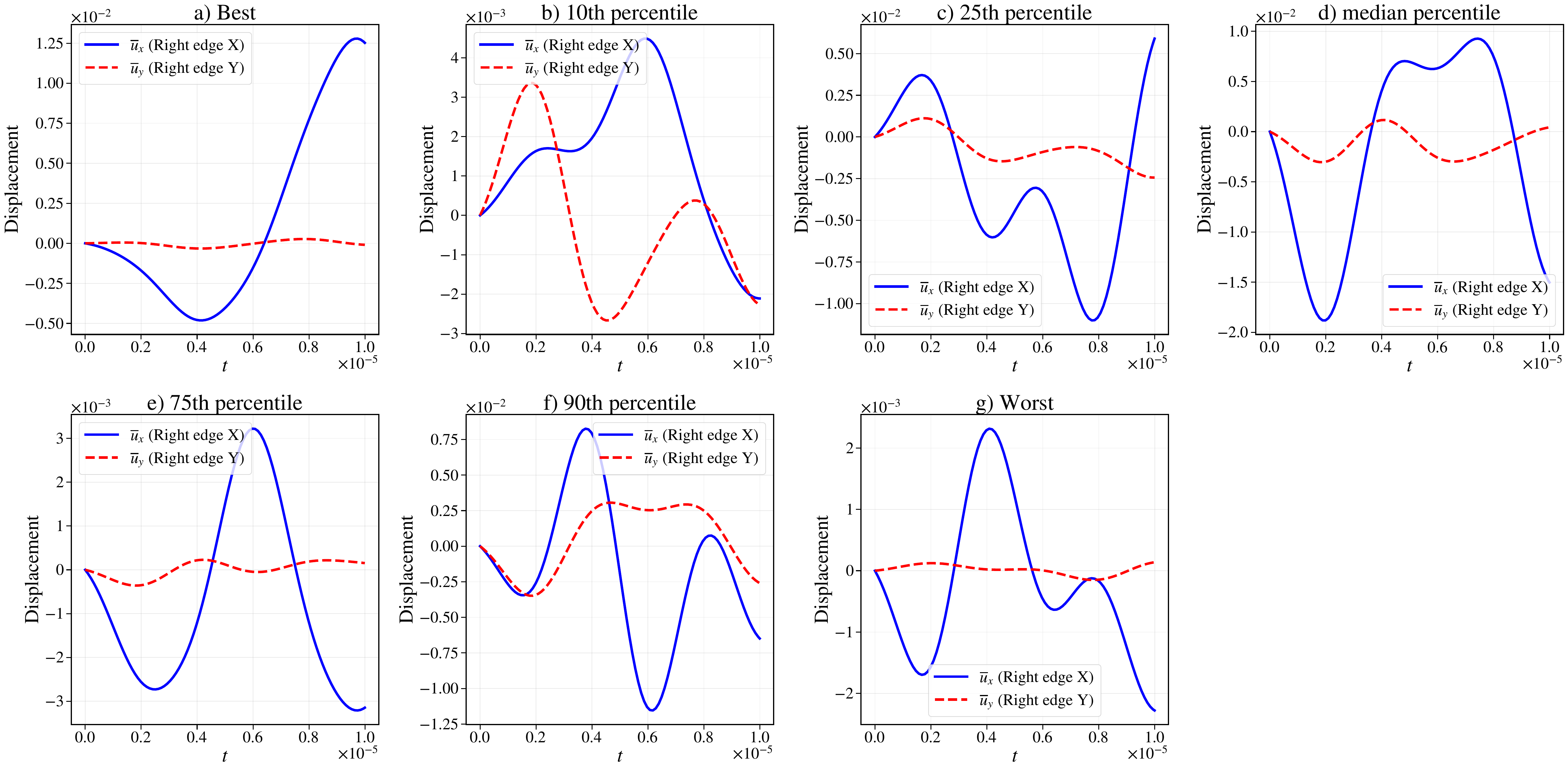}
    \caption{Input displacement boundary conditions $\bar{u}_x(t)$ and $\bar{u}_y(t)$ for elastodynamics test cases at different error percentiles. Each subplot shows the prescribed time-varying displacements applied at the right edge of the domain.}
    \label{fig:elastodynamics_input_signals}
\end{figure}

This section provides supplementary visualizations for the elastodynamics problem discussed in Section \ref{elastodynamics}. Figure \ref{fig:elastodynamics_input_signals} displays the input displacement boundary conditions for test cases spanning the entire error spectrum, from best to worst performance. These signals exhibit diverse temporal characteristics, including varying frequencies, amplitudes, and phase relationships between $\bar{u}_x(t)$ and $\bar{u}_y(t)$, demonstrating the range of dynamic loading scenarios included in the test dataset. The prescribed displacements range from smooth variations to complex multi-frequency oscillations, challenging the model's ability to capture different wave propagation patterns in the elastic medium.

\begin{figure}[!htb]
    \centering
    \includegraphics[width=\textwidth]{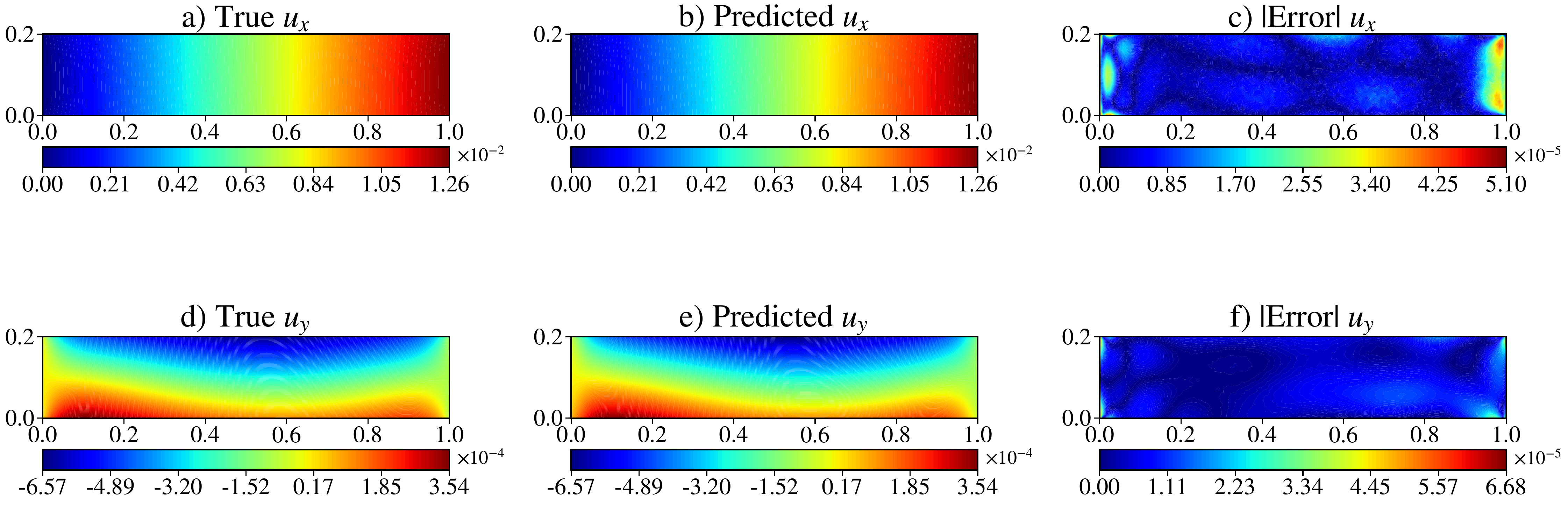}
    \caption{Elastodynamics field predictions at $t = 10^{-5}$s for the best test case. Columns show true fields, NCDE-DeepONet predictions, and absolute errors for displacement components $u_x$ and $u_y$.}
    \label{fig:elastodynamics_heatmap_best}
\end{figure}

\begin{figure}[!htb]
    \centering
    \includegraphics[width=\textwidth]{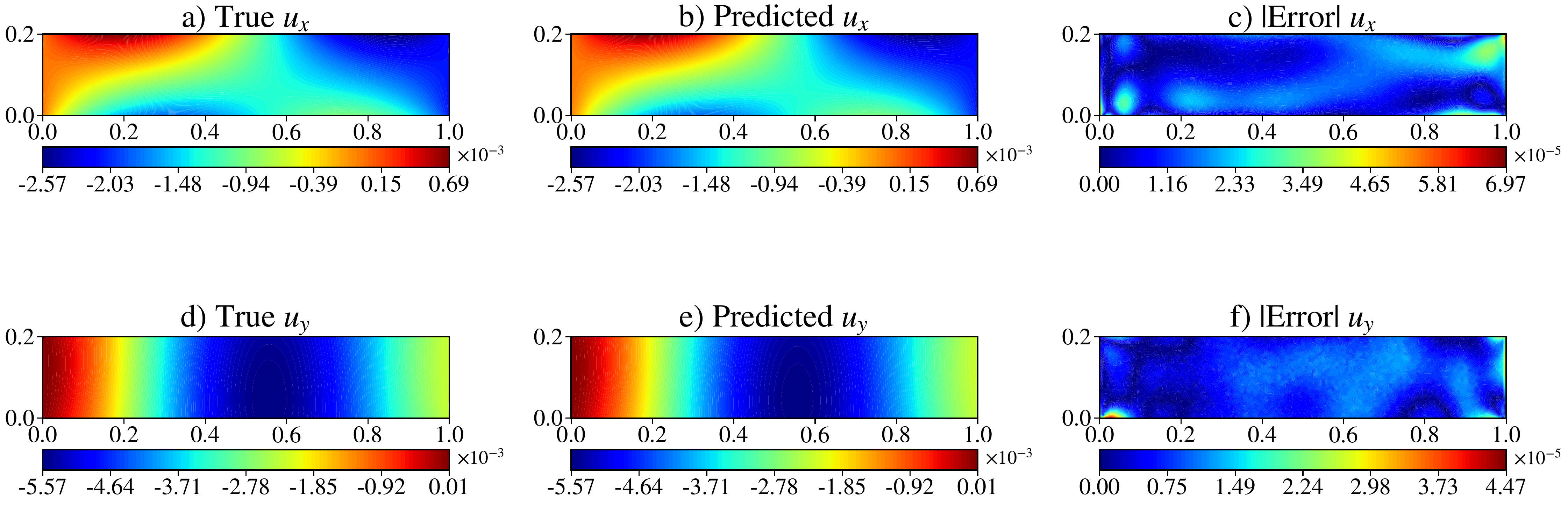}
    \caption{Elastodynamics field predictions at $t = 10^{-5}$s for the 10th percentile test case.}
    \label{fig:elastodynamics_heatmap_10th}
\end{figure}

\begin{figure}[!htb]
    \centering
    \includegraphics[width=\textwidth]{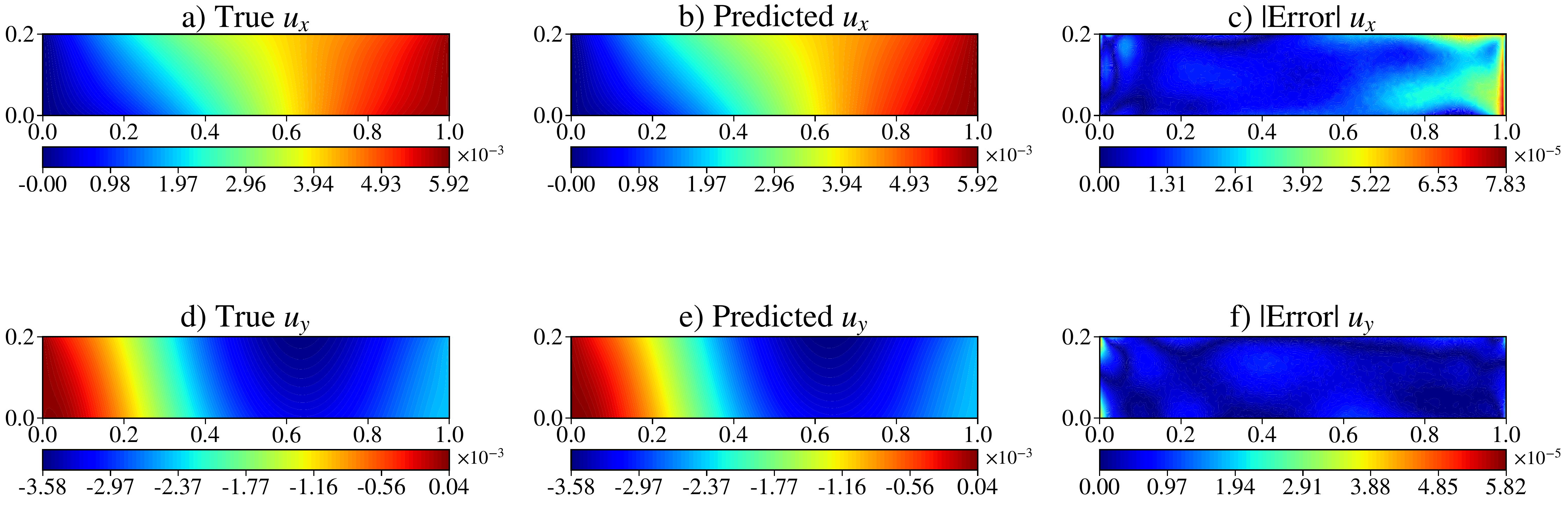}
    \caption{Elastodynamics field predictions at $t = 10^{-5}$s for the 25th percentile test case.}
    \label{fig:elastodynamics_heatmap_25th}
\end{figure}

\begin{figure}[!htb]
    \centering
    \includegraphics[width=\textwidth]{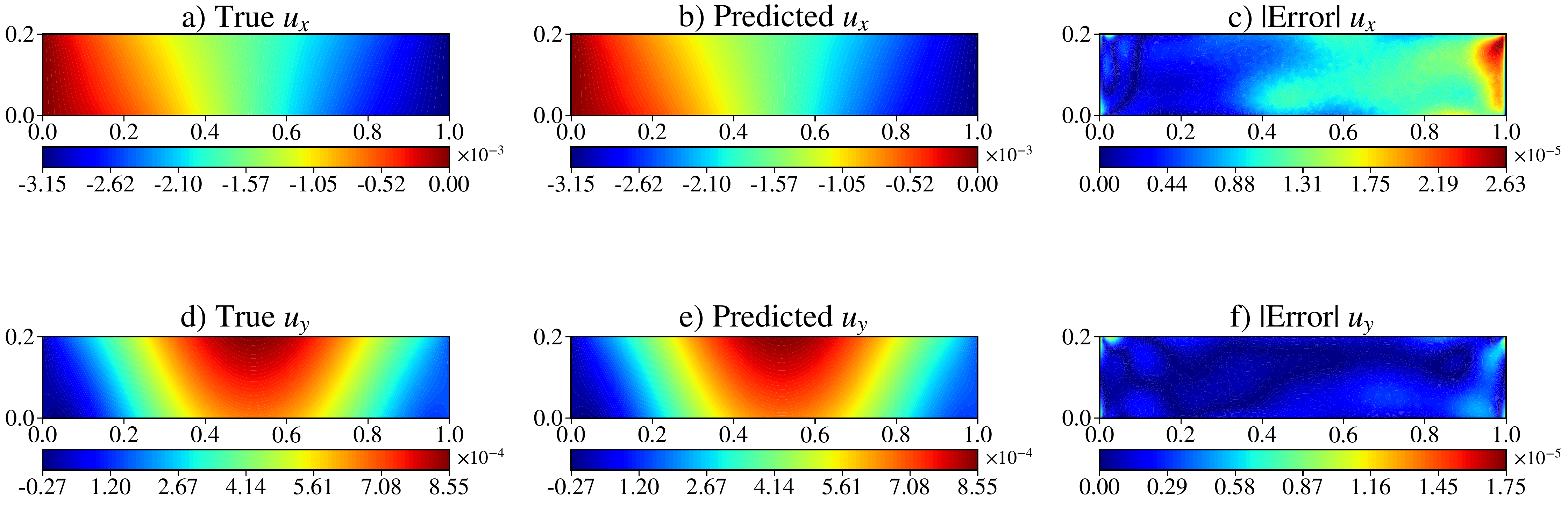}
    \caption{Elastodynamics field predictions at $t = 10^{-5}$s for the 75th percentile test case.}
    \label{fig:elastodynamics_heatmap_75th}
\end{figure}

\begin{figure}[!htb]
    \centering
    \includegraphics[width=\textwidth]{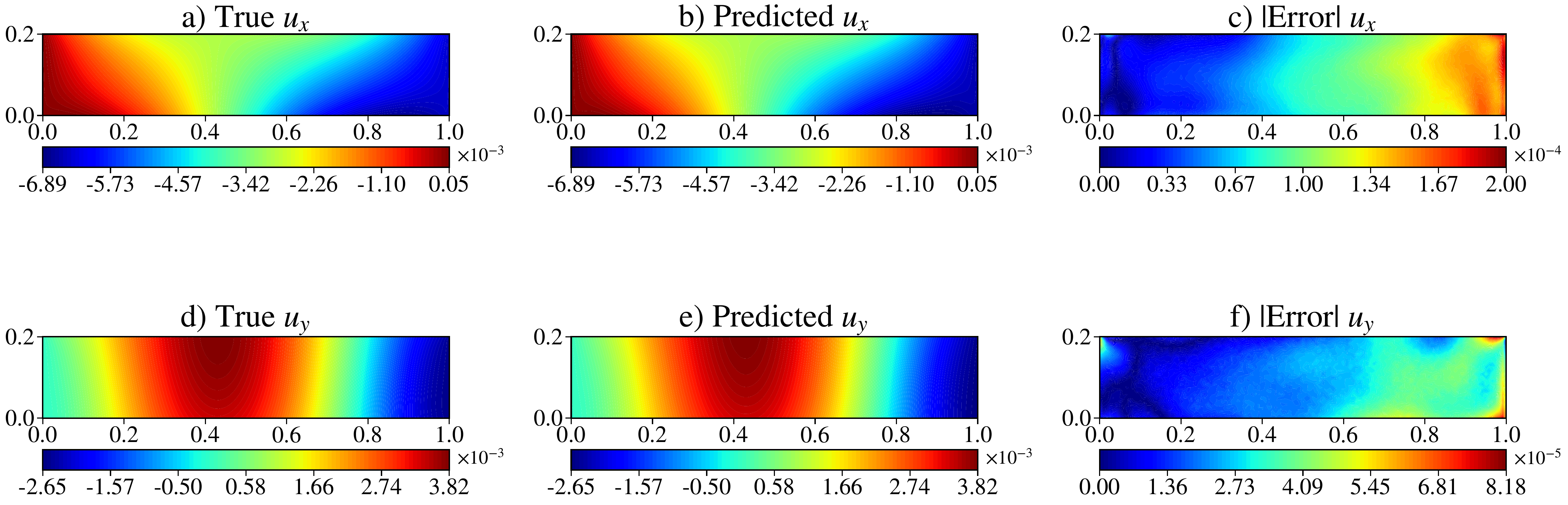}
    \caption{Elastodynamics field predictions at $t = 10^{-5}$s for the 90th percentile test case.}
    \label{fig:elastodynamics_heatmap_90th}
\end{figure}

\begin{figure}[!htb]
    \centering
    \includegraphics[width=\textwidth]{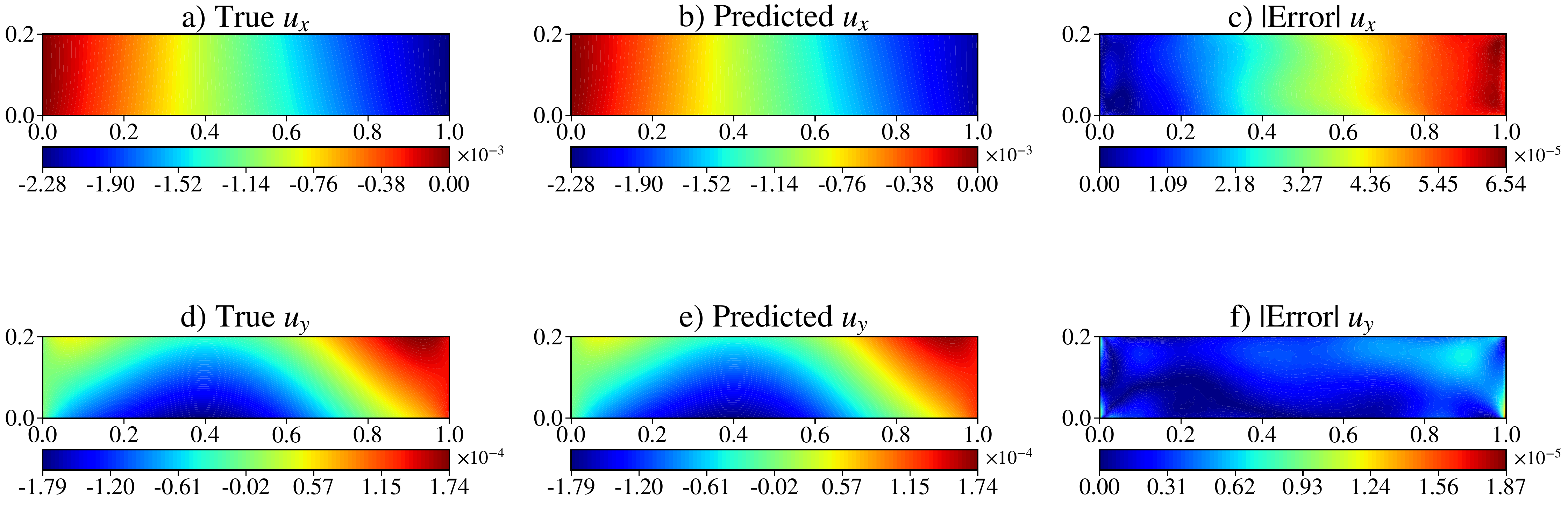}
    \caption{Elastodynamics field predictions at $t = 10^{-5}$s for the worst test case.}
    \label{fig:elastodynamics_heatmap_worst}
\end{figure}

Figures \ref{fig:elastodynamics_heatmap_best}--\ref{fig:elastodynamics_heatmap_worst} present the spatial field distributions at the final time $t = 10^{-5}$s for test cases at different error percentiles. These visualizations reveal several important observations. First, the model maintains physical consistency across all error levels, accurately capturing the elastodynamic response including wave propagation and deformation patterns, even in the worst-performing cases. Second, the error patterns remain relatively consistent across percentiles, with larger discrepancies typically occurring near the right boundary where displacements are prescribed and in regions of high stress concentration. Finally, comparing the best and worst cases shows that prediction errors are primarily related to the complexity and frequency content of the input displacement signals rather than any systematic model deficiency, as evidenced by the model's ability to reproduce the characteristic wave patterns and coupled displacement fields across all test cases.

\subsection*{A.4. Additional results: Thermoelasticity}

\begin{figure}[!htb]
    \centering
    \includegraphics[width=\textwidth]{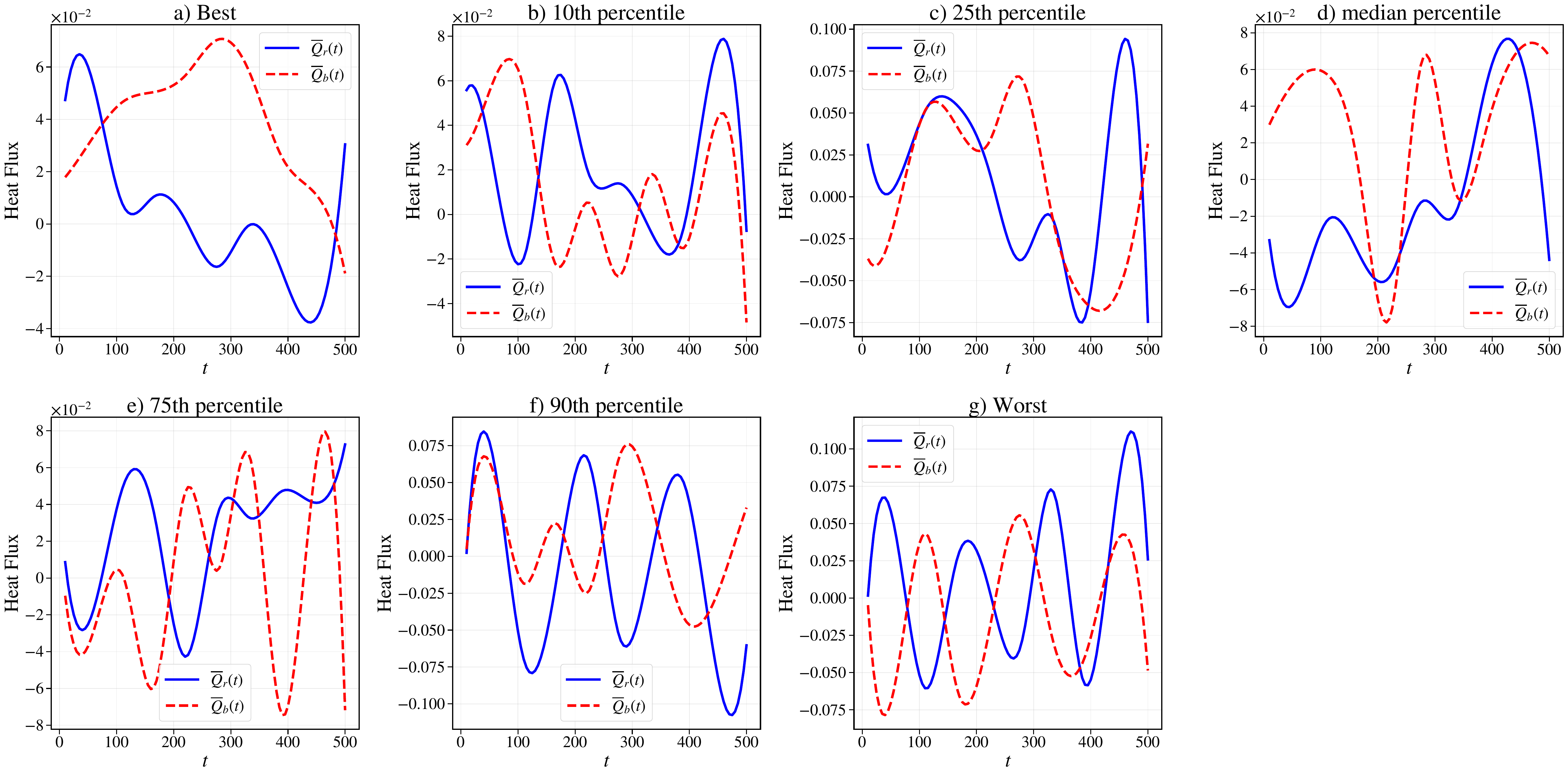}
    \caption{Input heat flux boundary conditions $\bar{Q}_r(t)$ (solid blue) and $\bar{Q}_b(t)$ (dashed red) for test cases at different error percentiles. The diverse signal characteristics demonstrate the model's ability to handle various loading scenarios with consistent accuracy.}
    \label{fig:thermoelasticity_input_signals}
\end{figure}

This section provides supplementary visualizations for the thermoelasticity problem discussed in Section \ref{thermoelasticity}. Figure \ref{fig:thermoelasticity_input_signals} displays the input heat flux boundary conditions for test cases spanning the entire error spectrum, from best to worst performance. These signals exhibit diverse temporal characteristics, including varying frequencies, amplitudes, and phase relationships between $\bar{Q}_r(t)$ and $\bar{Q}_b(t)$, demonstrating the range of loading scenarios included in the test dataset.

\begin{figure}[!htb]
    \centering
    \includegraphics[width=\textwidth]{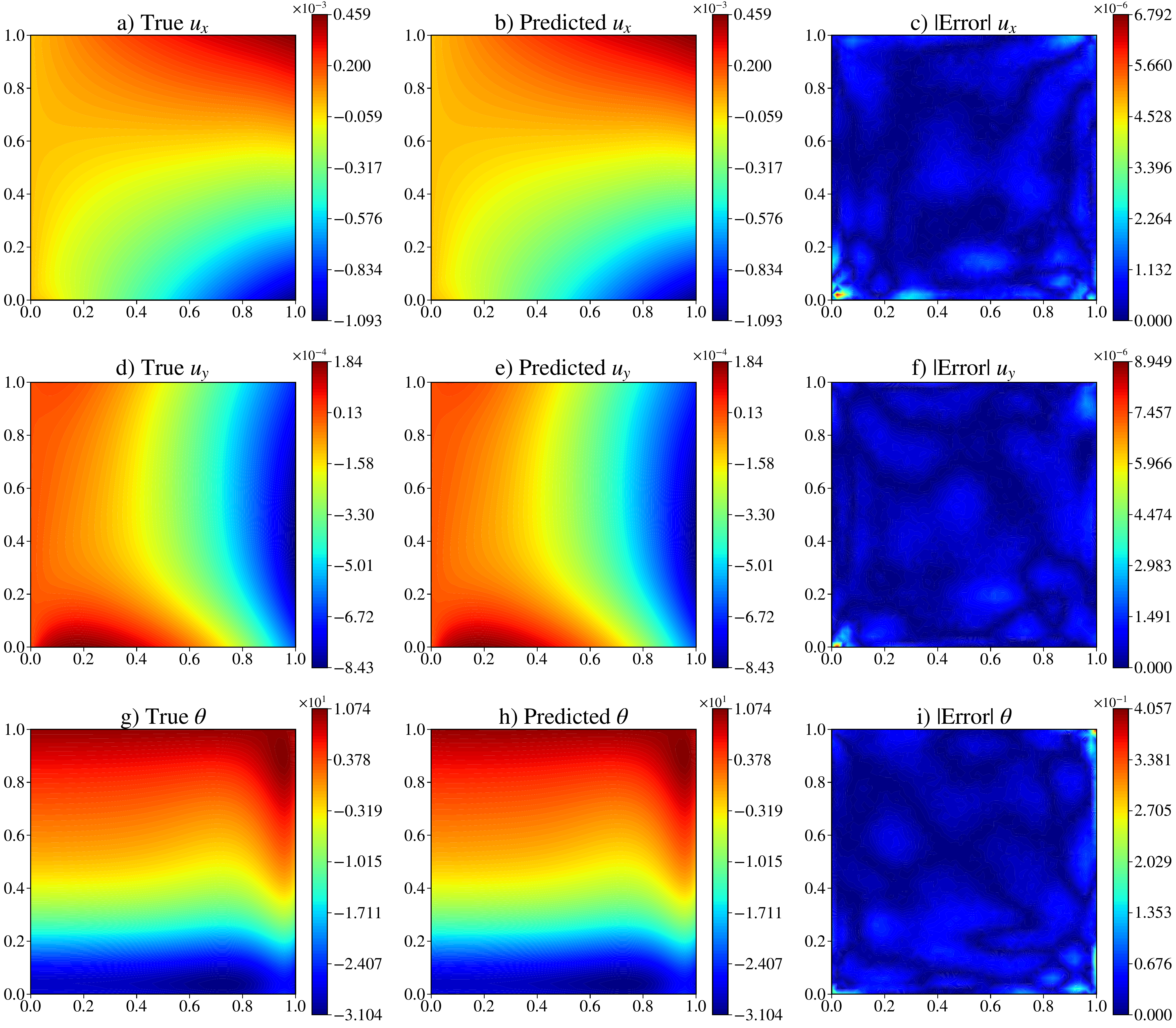}
    \caption{Thermoelasticity field predictions at $t = 500$s for the best test case. Columns show true fields, NCDE-DeepONet predictions, and absolute errors for displacement components $u_x$, $u_y$, and temperature $\theta$.}
    \label{fig:thermoelasticity_heatmap_best}
\end{figure}

\begin{figure}[!htb]
    \centering
    \includegraphics[width=\textwidth]{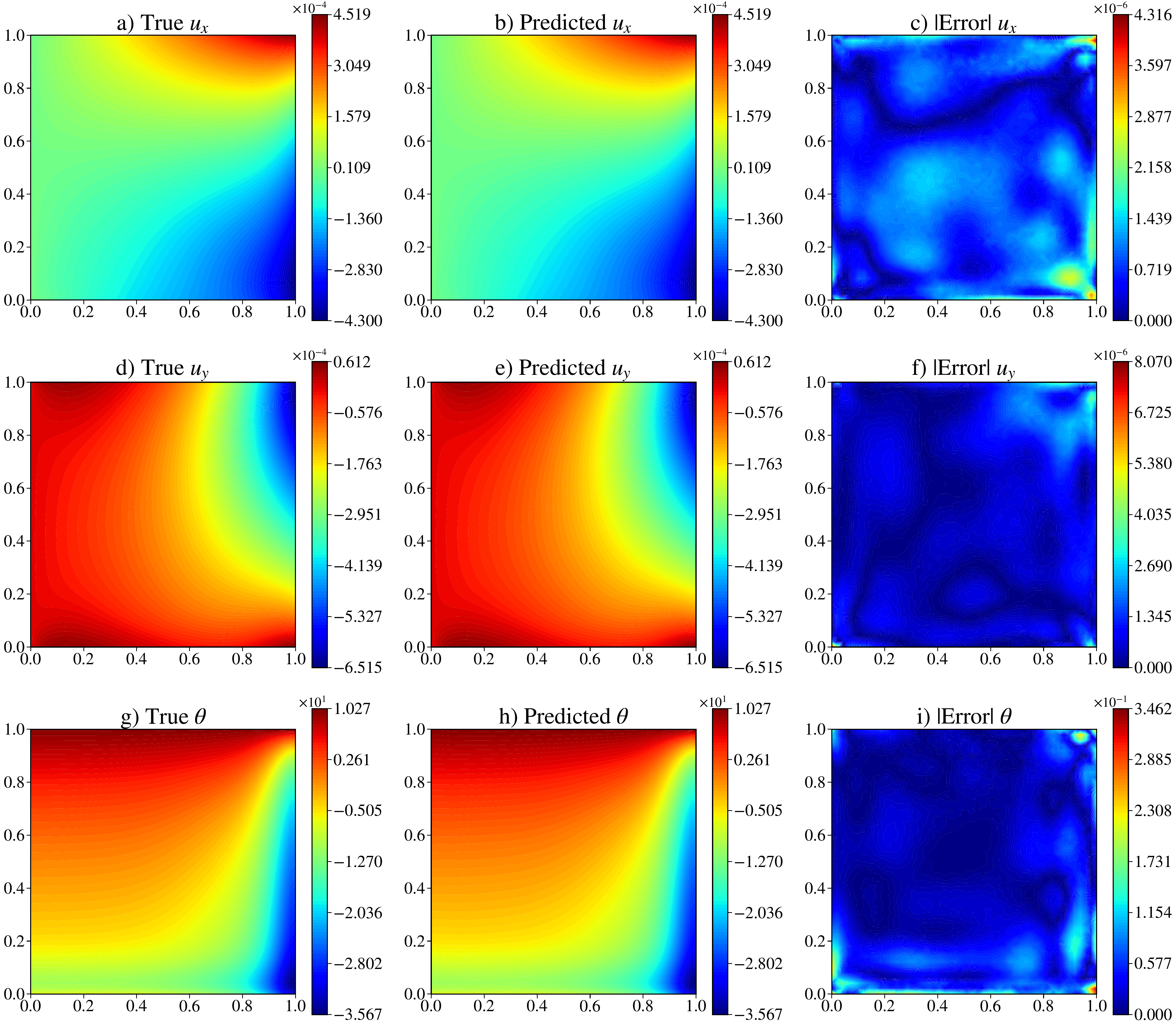}
    \caption{Thermoelasticity field predictions at $t = 500$s for the 10th percentile error test case.}
    \label{fig:thermoelasticity_heatmap_10th}
\end{figure}

\begin{figure}[!htb]
    \centering
    \includegraphics[width=\textwidth]{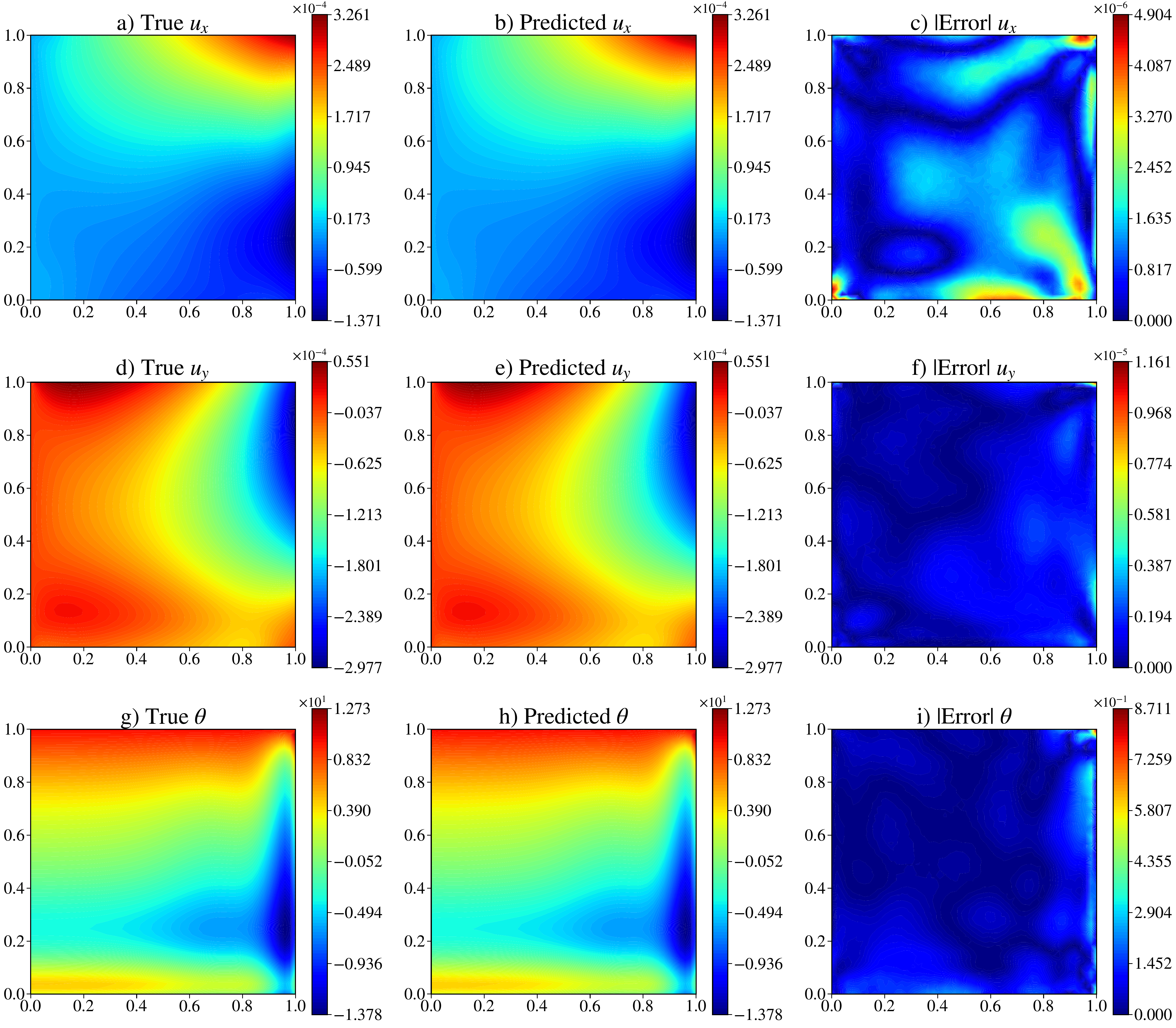}
    \caption{Thermoelasticity field predictions at $t = 500$s for the 25th percentile error test case.}
    \label{fig:thermoelasticity_heatmap_25th}
\end{figure}

\begin{figure}[!htb]
    \centering
    \includegraphics[width=\textwidth]{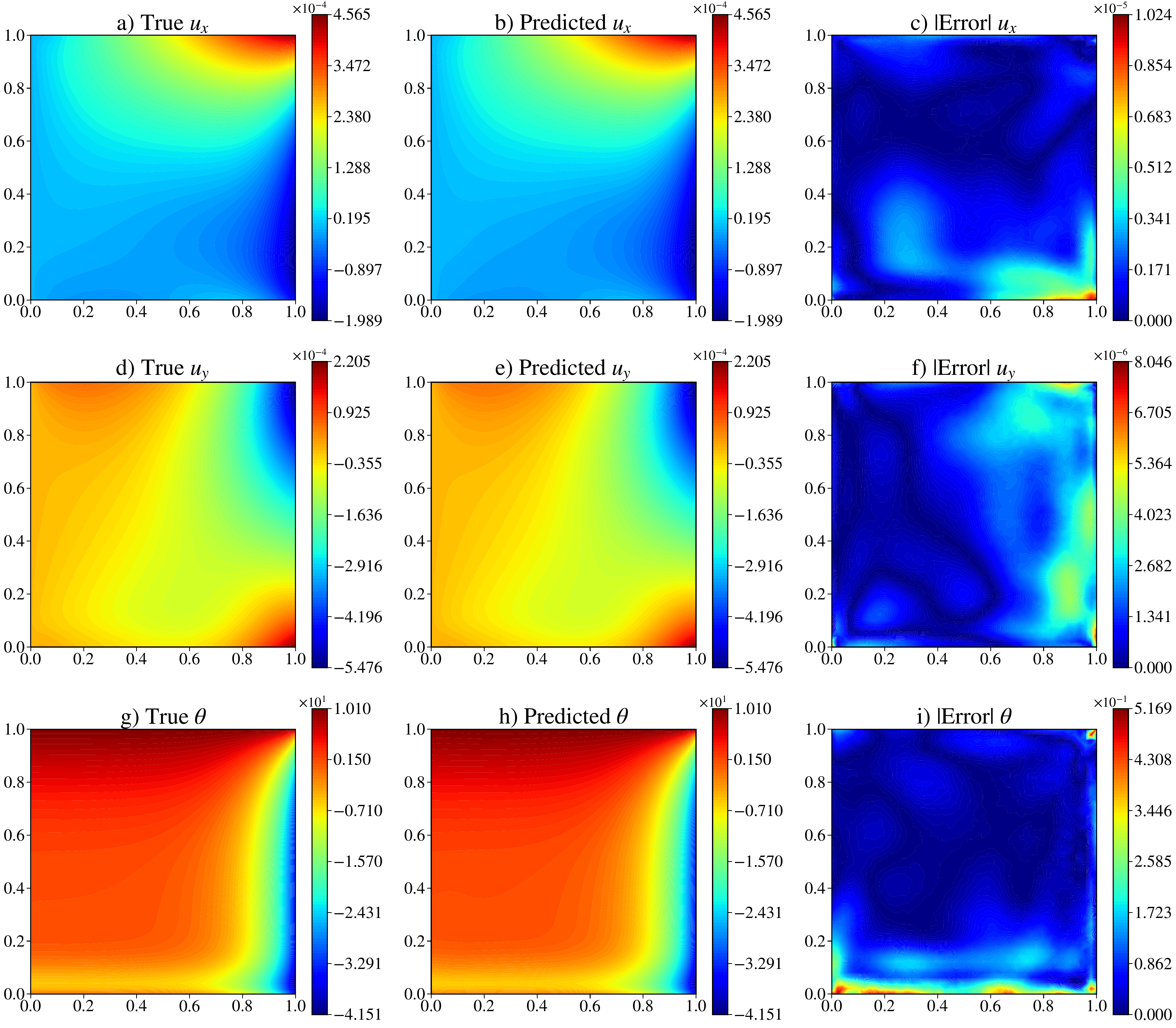}
    \caption{Thermoelasticity field predictions at $t = 500$s for the 75th percentile error test case.}
    \label{fig:thermoelasticity_heatmap_75th}
\end{figure}

\begin{figure}[!htb]
    \centering
    \includegraphics[width=\textwidth]{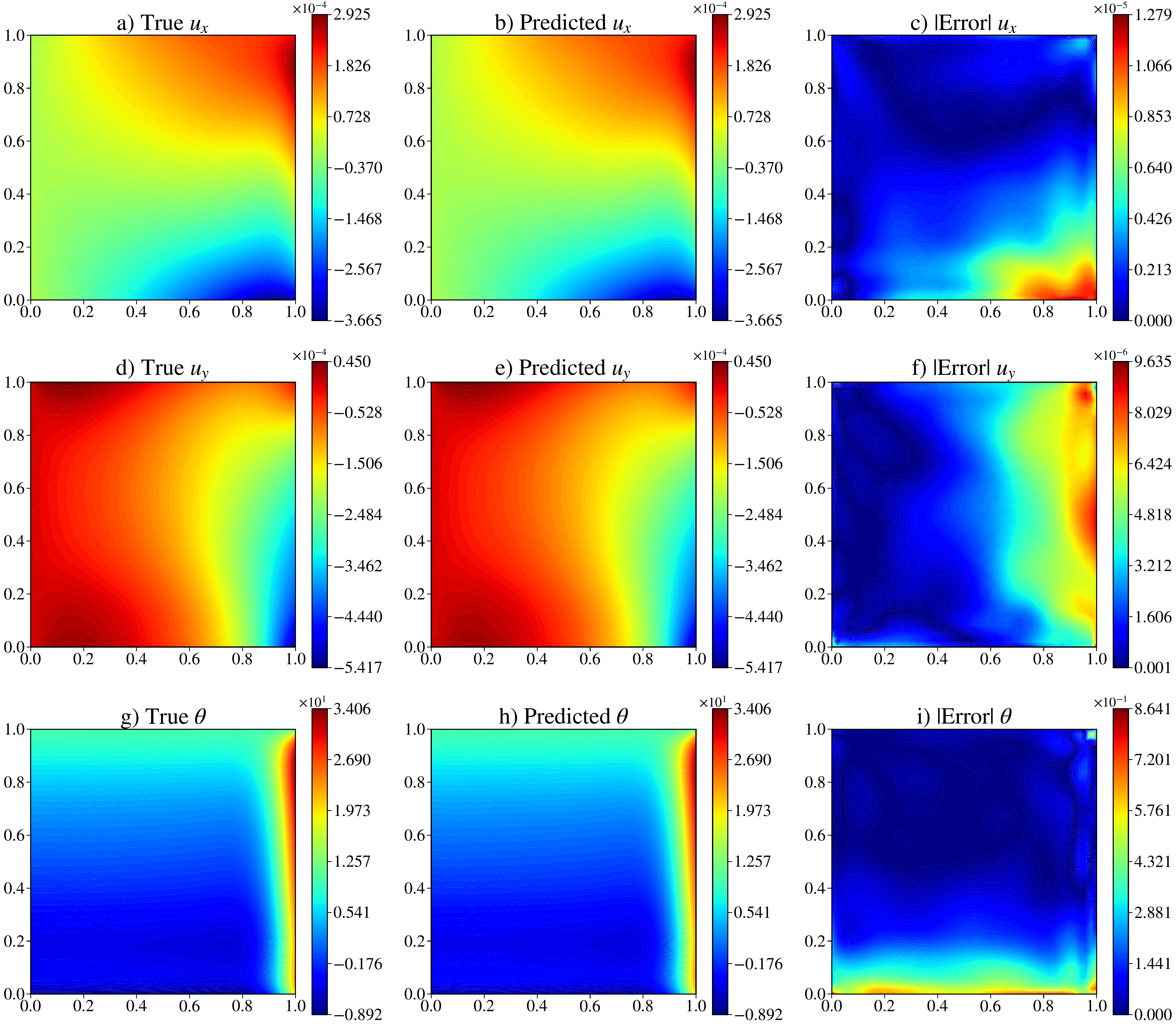}
    \caption{Thermoelasticity field predictions at $t = 500$s for the 90th percentile error test case.}
    \label{fig:thermoelasticity_heatmap_90th}
\end{figure}

\begin{figure}[!htb]
    \centering
    \includegraphics[width=\textwidth]{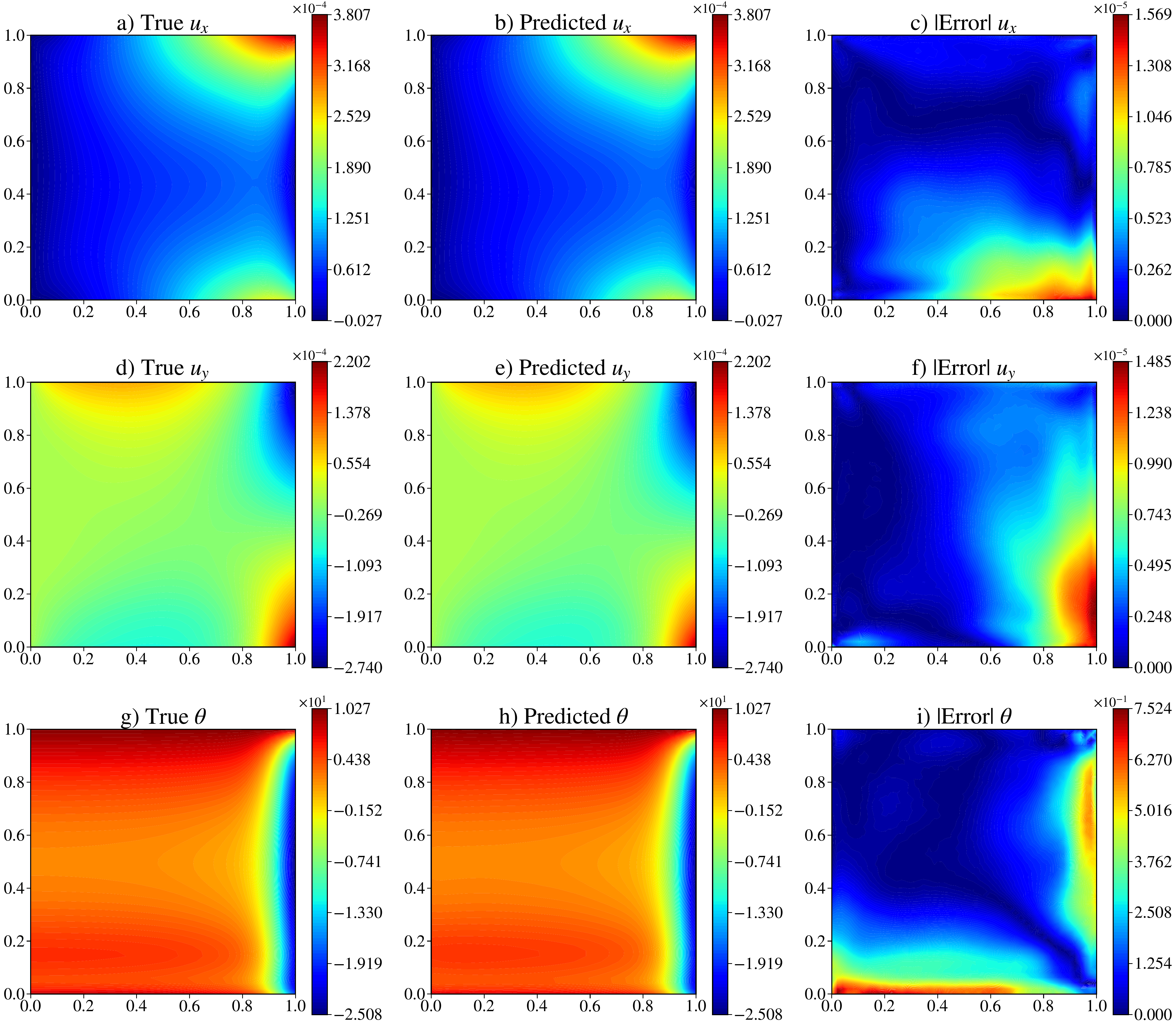}
    \caption{Thermoelasticity field predictions at $t = 500$s for the worst test case. Despite having the highest error, the model still captures the essential features of the thermoelastic response.}
    \label{fig:thermoelasticity_heatmap_worst}
\end{figure}

Figures \ref{fig:thermoelasticity_heatmap_best}--\ref{fig:thermoelasticity_heatmap_worst} present the spatial field distributions at the final time $t = 500$s for test cases at different error percentiles. These visualizations reveal several important observations. First, the model maintains physical consistency across all error levels, accurately capturing the coupled thermoelastic behavior, even in the worst-performing cases. Second, the error patterns remain relatively consistent across percentiles, with larger discrepancies typically occurring near the flux boundaries and regions of high gradient. Finally, comparing the best and worst cases shows that prediction errors are primarily related to the complexity of the input signals rather than any systematic model deficiency, as evidenced by the model's ability to reproduce the qualitative field patterns across all test cases.

\end{document}